
\catcode`\@=11


\message{Loading jyTeX fonts...}



\font\vptrm=cmr5 \font\vptmit=cmmi5 \font\vptsy=cmsy5 \font\vptbf=cmbx5

\skewchar\vptmit='177 \skewchar\vptsy='60 \fontdimen16 \vptsy=\the\fontdimen17 \vptsy

\def\vpt{\ifmmode\err@badsizechange\else
     \@mathfontinit
     \textfont0=\vptrm  \scriptfont0=\vptrm  \scriptscriptfont0=\vptrm
     \textfont1=\vptmit \scriptfont1=\vptmit \scriptscriptfont1=\vptmit
     \textfont2=\vptsy  \scriptfont2=\vptsy  \scriptscriptfont2=\vptsy
     \textfont3=\xptex  \scriptfont3=\xptex  \scriptscriptfont3=\xptex
     \textfont\bffam=\vptbf
     \scriptfont\bffam=\vptbf
     \scriptscriptfont\bffam=\vptbf
     \@fontstyleinit
     \def\rm{\vptrm\fam=\z@}%
     \def\bf{\vptbf\fam=\bffam}%
     \def\oldstyle{\vptmit\fam=\@ne}%
     \rm\fi}


\font\viptrm=cmr6 \font\viptmit=cmmi6 \font\viptsy=cmsy6 \font\viptbf=cmbx6

\skewchar\viptmit='177 \skewchar\viptsy='60 \fontdimen16 \viptsy=\the\fontdimen17
\viptsy

\def\vipt{\ifmmode\err@badsizechange\else
     \@mathfontinit
     \textfont0=\viptrm  \scriptfont0=\vptrm  \scriptscriptfont0=\vptrm
     \textfont1=\viptmit \scriptfont1=\vptmit \scriptscriptfont1=\vptmit
     \textfont2=\viptsy  \scriptfont2=\vptsy  \scriptscriptfont2=\vptsy
     \textfont3=\xptex   \scriptfont3=\xptex  \scriptscriptfont3=\xptex
     \textfont\bffam=\viptbf
     \scriptfont\bffam=\vptbf
     \scriptscriptfont\bffam=\vptbf
     \@fontstyleinit
     \def\rm{\viptrm\fam=\z@}%
     \def\bf{\viptbf\fam=\bffam}%
     \def\oldstyle{\viptmit\fam=\@ne}%
     \rm\fi}

\font\viiptrm=cmr7 \font\viiptmit=cmmi7 \font\viiptsy=cmsy7 \font\viiptit=cmti7
\font\viiptbf=cmbx7

\skewchar\viiptmit='177 \skewchar\viiptsy='60 \fontdimen16 \viiptsy=\the\fontdimen17
\viiptsy

\def\viipt{\ifmmode\err@badsizechange\else
     \@mathfontinit
     \textfont0=\viiptrm  \scriptfont0=\vptrm  \scriptscriptfont0=\vptrm
     \textfont1=\viiptmit \scriptfont1=\vptmit \scriptscriptfont1=\vptmit
     \textfont2=\viiptsy  \scriptfont2=\vptsy  \scriptscriptfont2=\vptsy
     \textfont3=\xptex    \scriptfont3=\xptex  \scriptscriptfont3=\xptex
     \textfont\itfam=\viiptit
     \scriptfont\itfam=\viiptit
     \scriptscriptfont\itfam=\viiptit
     \textfont\bffam=\viiptbf
     \scriptfont\bffam=\vptbf
     \scriptscriptfont\bffam=\vptbf
     \@fontstyleinit
     \def\rm{\viiptrm\fam=\z@}%
     \def\it{\viiptit\fam=\itfam}%
     \def\bf{\viiptbf\fam=\bffam}%
     \def\oldstyle{\viiptmit\fam=\@ne}%
     \rm\fi}


\font\viiiptrm=cmr8 \font\viiiptmit=cmmi8 \font\viiiptsy=cmsy8 \font\viiiptit=cmti8
\font\viiiptbf=cmbx8

\skewchar\viiiptmit='177 \skewchar\viiiptsy='60 \fontdimen16 \viiiptsy=\the\fontdimen17
\viiiptsy

\def\viiipt{\ifmmode\err@badsizechange\else
     \@mathfontinit
     \textfont0=\viiiptrm  \scriptfont0=\viptrm  \scriptscriptfont0=\vptrm
     \textfont1=\viiiptmit \scriptfont1=\viptmit \scriptscriptfont1=\vptmit
     \textfont2=\viiiptsy  \scriptfont2=\viptsy  \scriptscriptfont2=\vptsy
     \textfont3=\xptex     \scriptfont3=\xptex   \scriptscriptfont3=\xptex
     \textfont\itfam=\viiiptit
     \scriptfont\itfam=\viiptit
     \scriptscriptfont\itfam=\viiptit
     \textfont\bffam=\viiiptbf
     \scriptfont\bffam=\viptbf
     \scriptscriptfont\bffam=\vptbf
     \@fontstyleinit
     \def\rm{\viiiptrm\fam=\z@}%
     \def\it{\viiiptit\fam=\itfam}%
     \def\bf{\viiiptbf\fam=\bffam}%
     \def\oldstyle{\viiiptmit\fam=\@ne}%
     \rm\fi}


\def\getixpt{%
     \font\ixptrm=cmr9
     \font\ixptmit=cmmi9
     \font\ixptsy=cmsy9
     \font\ixptit=cmti9
     \font\ixptbf=cmbx9
     \skewchar\ixptmit='177 \skewchar\ixptsy='60
     \fontdimen16 \ixptsy=\the\fontdimen17 \ixptsy}

\def\ixpt{\ifmmode\err@badsizechange\else
     \@mathfontinit
     \textfont0=\ixptrm  \scriptfont0=\viiptrm  \scriptscriptfont0=\vptrm
     \textfont1=\ixptmit \scriptfont1=\viiptmit \scriptscriptfont1=\vptmit
     \textfont2=\ixptsy  \scriptfont2=\viiptsy  \scriptscriptfont2=\vptsy
     \textfont3=\xptex   \scriptfont3=\xptex    \scriptscriptfont3=\xptex
     \textfont\itfam=\ixptit
     \scriptfont\itfam=\viiptit
     \scriptscriptfont\itfam=\viiptit
     \textfont\bffam=\ixptbf
     \scriptfont\bffam=\viiptbf
     \scriptscriptfont\bffam=\vptbf
     \@fontstyleinit
     \def\rm{\ixptrm\fam=\z@}%
     \def\it{\ixptit\fam=\itfam}%
     \def\bf{\ixptbf\fam=\bffam}%
     \def\oldstyle{\ixptmit\fam=\@ne}%
     \rm\fi}


\font\xptrm=cmr10 \font\xptmit=cmmi10 \font\xptsy=cmsy10 \font\xptex=cmex10
\font\xptit=cmti10 \font\xptsl=cmsl10 \font\xptbf=cmbx10 \font\xpttt=cmtt10
\font\xptss=cmss10 \font\xptsc=cmcsc10 \font\xptbfs=cmb10 \font\xptbmit=cmmib10

\skewchar\xptmit='177 \skewchar\xptbmit='177 \skewchar\xptsy='60 \fontdimen16
\xptsy=\the\fontdimen17 \xptsy

\def\xpt{\ifmmode\err@badsizechange\else
     \@mathfontinit
     \textfont0=\xptrm  \scriptfont0=\viiptrm  \scriptscriptfont0=\vptrm
     \textfont1=\xptmit \scriptfont1=\viiptmit \scriptscriptfont1=\vptmit
     \textfont2=\xptsy  \scriptfont2=\viiptsy  \scriptscriptfont2=\vptsy
     \textfont3=\xptex  \scriptfont3=\xptex    \scriptscriptfont3=\xptex
     \textfont\itfam=\xptit
     \scriptfont\itfam=\viiptit
     \scriptscriptfont\itfam=\viiptit
     \textfont\bffam=\xptbf
     \scriptfont\bffam=\viiptbf
     \scriptscriptfont\bffam=\vptbf
     \textfont\bfsfam=\xptbfs
     \scriptfont\bfsfam=\viiptbf
     \scriptscriptfont\bfsfam=\vptbf
     \textfont\bmitfam=\xptbmit
     \scriptfont\bmitfam=\viiptmit
     \scriptscriptfont\bmitfam=\vptmit
     \@fontstyleinit
     \def\rm{\xptrm\fam=\z@}%
     \def\it{\xptit\fam=\itfam}%
     \def\sl{\xptsl}%
     \def\bf{\xptbf\fam=\bffam}%
     \def\tt{\xpttt}%
     \def\ss{\xptss}%
     \def\sc{\xptsc}%
     \def\bfs{\xptbfs\fam=\bfsfam}%
     \def\bmit{\fam=\bmitfam}%
     \def\oldstyle{\xptmit\fam=\@ne}%
     \rm\fi}


\def\getxipt{%
     \font\xiptrm=cmr10  scaled\magstephalf
     \font\xiptmit=cmmi10 scaled\magstephalf
     \font\xiptsy=cmsy10 scaled\magstephalf
     \font\xiptex=cmex10 scaled\magstephalf
     \font\xiptit=cmti10 scaled\magstephalf
     \font\xiptsl=cmsl10 scaled\magstephalf
     \font\xiptbf=cmbx10 scaled\magstephalf
     \font\xipttt=cmtt10 scaled\magstephalf
     \font\xiptss=cmss10 scaled\magstephalf
     \skewchar\xiptmit='177 \skewchar\xiptsy='60
     \fontdimen16 \xiptsy=\the\fontdimen17 \xiptsy}

\def\xipt{\ifmmode\err@badsizechange\else
     \@mathfontinit
     \textfont0=\xiptrm  \scriptfont0=\viiiptrm  \scriptscriptfont0=\viptrm
     \textfont1=\xiptmit \scriptfont1=\viiiptmit \scriptscriptfont1=\viptmit
     \textfont2=\xiptsy  \scriptfont2=\viiiptsy  \scriptscriptfont2=\viptsy
     \textfont3=\xiptex  \scriptfont3=\xptex     \scriptscriptfont3=\xptex
     \textfont\itfam=\xiptit
     \scriptfont\itfam=\viiiptit
     \scriptscriptfont\itfam=\viiptit
     \textfont\bffam=\xiptbf
     \scriptfont\bffam=\viiiptbf
     \scriptscriptfont\bffam=\viptbf
     \@fontstyleinit
     \def\rm{\xiptrm\fam=\z@}%
     \def\it{\xiptit\fam=\itfam}%
     \def\sl{\xiptsl}%
     \def\bf{\xiptbf\fam=\bffam}%
     \def\tt{\xipttt}%
     \def\ss{\xiptss}%
     \def\oldstyle{\xiptmit\fam=\@ne}%
     \rm\fi}


\font\xiiptrm=cmr12 \font\xiiptmit=cmmi12 \font\xiiptsy=cmsy10 scaled\magstep1
\font\xiiptex=cmex10  scaled\magstep1 \font\xiiptit=cmti12 \font\xiiptsl=cmsl12
\font\xiiptbf=cmbx12
\font\xiiptss=cmss12 \font\xiiptsc=cmcsc10 scaled\magstep1 \font\xiiptbfs=cmb10
scaled\magstep1 \font\xiiptbmit=cmmib10 scaled\magstep1

\skewchar\xiiptmit='177 \skewchar\xiiptbmit='177 \skewchar\xiiptsy='60 \fontdimen16
\xiiptsy=\the\fontdimen17 \xiiptsy

\def\xiipt{\ifmmode\err@badsizechange\else
     \@mathfontinit
     \textfont0=\xiiptrm  \scriptfont0=\viiiptrm  \scriptscriptfont0=\viptrm
     \textfont1=\xiiptmit \scriptfont1=\viiiptmit \scriptscriptfont1=\viptmit
     \textfont2=\xiiptsy  \scriptfont2=\viiiptsy  \scriptscriptfont2=\viptsy
     \textfont3=\xiiptex  \scriptfont3=\xptex     \scriptscriptfont3=\xptex
     \textfont\itfam=\xiiptit
     \scriptfont\itfam=\viiiptit
     \scriptscriptfont\itfam=\viiptit
     \textfont\bffam=\xiiptbf
     \scriptfont\bffam=\viiiptbf
     \scriptscriptfont\bffam=\viptbf
     \textfont\bfsfam=\xiiptbfs
     \scriptfont\bfsfam=\viiiptbf
     \scriptscriptfont\bfsfam=\viptbf
     \textfont\bmitfam=\xiiptbmit
     \scriptfont\bmitfam=\viiiptmit
     \scriptscriptfont\bmitfam=\viptmit
     \@fontstyleinit
     \def\rm{\xiiptrm\fam=\z@}%
     \def\it{\xiiptit\fam=\itfam}%
     \def\sl{\xiiptsl}%
     \def\bf{\xiiptbf\fam=\bffam}%
     \def\tt{\xiipttt}%
     \def\ss{\xiiptss}%
     \def\sc{\xiiptsc}%
     \def\bfs{\xiiptbfs\fam=\bfsfam}%
     \def\bmit{\fam=\bmitfam}%
     \def\oldstyle{\xiiptmit\fam=\@ne}%
     \rm\fi}


\def\getxiiipt{%
     \font\xiiiptrm=cmr12  scaled\magstephalf
     \font\xiiiptmit=cmmi12 scaled\magstephalf
     \font\xiiiptsy=cmsy9  scaled\magstep2
     \font\xiiiptit=cmti12 scaled\magstephalf
     \font\xiiiptsl=cmsl12 scaled\magstephalf
     \font\xiiiptbf=cmbx12 scaled\magstephalf
     \font\xiiipttt=cmtt12 scaled\magstephalf
     \font\xiiiptss=cmss12 scaled\magstephalf
     \skewchar\xiiiptmit='177 \skewchar\xiiiptsy='60
     \fontdimen16 \xiiiptsy=\the\fontdimen17 \xiiiptsy}

\def\xiiipt{\ifmmode\err@badsizechange\else
     \@mathfontinit
     \textfont0=\xiiiptrm  \scriptfont0=\xptrm  \scriptscriptfont0=\viiptrm
     \textfont1=\xiiiptmit \scriptfont1=\xptmit \scriptscriptfont1=\viiptmit
     \textfont2=\xiiiptsy  \scriptfont2=\xptsy  \scriptscriptfont2=\viiptsy
     \textfont3=\xivptex   \scriptfont3=\xptex  \scriptscriptfont3=\xptex
     \textfont\itfam=\xiiiptit
     \scriptfont\itfam=\xptit
     \scriptscriptfont\itfam=\viiptit
     \textfont\bffam=\xiiiptbf
     \scriptfont\bffam=\xptbf
     \scriptscriptfont\bffam=\viiptbf
     \@fontstyleinit
     \def\rm{\xiiiptrm\fam=\z@}%
     \def\it{\xiiiptit\fam=\itfam}%
     \def\sl{\xiiiptsl}%
     \def\bf{\xiiiptbf\fam=\bffam}%
     \def\tt{\xiiipttt}%
     \def\ss{\xiiiptss}%
     \def\oldstyle{\xiiiptmit\fam=\@ne}%
     \rm\fi}


\font\xivptrm=cmr12   scaled\magstep1 \font\xivptmit=cmmi12 scaled\magstep1
\font\xivptsy=cmsy10  scaled\magstep2 \font\xivptex=cmex10  scaled\magstep2
\font\xivptit=cmti12 scaled\magstep1 \font\xivptsl=cmsl12  scaled\magstep1
\font\xivptbf=cmbx12  scaled\magstep1
\font\xivptss=cmss12  scaled\magstep1 \font\xivptsc=cmcsc10 scaled\magstep2
\font\xivptbfs=cmb10  scaled\magstep2 \font\xivptbmit=cmmib10 scaled\magstep2

\skewchar\xivptmit='177 \skewchar\xivptbmit='177 \skewchar\xivptsy='60 \fontdimen16
\xivptsy=\the\fontdimen17 \xivptsy

\def\xivpt{\ifmmode\err@badsizechange\else
     \@mathfontinit
     \textfont0=\xivptrm  \scriptfont0=\xptrm  \scriptscriptfont0=\viiptrm
     \textfont1=\xivptmit \scriptfont1=\xptmit \scriptscriptfont1=\viiptmit
     \textfont2=\xivptsy  \scriptfont2=\xptsy  \scriptscriptfont2=\viiptsy
     \textfont3=\xivptex  \scriptfont3=\xptex  \scriptscriptfont3=\xptex
     \textfont\itfam=\xivptit
     \scriptfont\itfam=\xptit
     \scriptscriptfont\itfam=\viiptit
     \textfont\bffam=\xivptbf
     \scriptfont\bffam=\xptbf
     \scriptscriptfont\bffam=\viiptbf
     \textfont\bfsfam=\xivptbfs
     \scriptfont\bfsfam=\xptbfs
     \scriptscriptfont\bfsfam=\viiptbf
     \textfont\bmitfam=\xivptbmit
     \scriptfont\bmitfam=\xptbmit
     \scriptscriptfont\bmitfam=\viiptmit
     \@fontstyleinit
     \def\rm{\xivptrm\fam=\z@}%
     \def\it{\xivptit\fam=\itfam}%
     \def\sl{\xivptsl}%
     \def\bf{\xivptbf\fam=\bffam}%
     \def\tt{\xivpttt}%
     \def\ss{\xivptss}%
     \def\sc{\xivptsc}%
     \def\bfs{\xivptbfs\fam=\bfsfam}%
     \def\bmit{\fam=\bmitfam}%
     \def\oldstyle{\xivptmit\fam=\@ne}%
     \rm\fi}


\font\xviiptrm=cmr17 \font\xviiptmit=cmmi12 scaled\magstep2 \font\xviiptsy=cmsy10
scaled\magstep3 \font\xviiptex=cmex10 scaled\magstep3 \font\xviiptit=cmti12
scaled\magstep2 \font\xviiptbf=cmbx12 scaled\magstep2 \font\xviiptbfs=cmb10
scaled\magstep3

\skewchar\xviiptmit='177 \skewchar\xviiptsy='60 \fontdimen16 \xviiptsy=\the\fontdimen17
\xviiptsy

\def\xviipt{\ifmmode\err@badsizechange\else
     \@mathfontinit
     \textfont0=\xviiptrm  \scriptfont0=\xiiptrm  \scriptscriptfont0=\viiiptrm
     \textfont1=\xviiptmit \scriptfont1=\xiiptmit \scriptscriptfont1=\viiiptmit
     \textfont2=\xviiptsy  \scriptfont2=\xiiptsy  \scriptscriptfont2=\viiiptsy
     \textfont3=\xviiptex  \scriptfont3=\xiiptex  \scriptscriptfont3=\xptex
     \textfont\itfam=\xviiptit
     \scriptfont\itfam=\xiiptit
     \scriptscriptfont\itfam=\viiiptit
     \textfont\bffam=\xviiptbf
     \scriptfont\bffam=\xiiptbf
     \scriptscriptfont\bffam=\viiiptbf
     \textfont\bfsfam=\xviiptbfs
     \scriptfont\bfsfam=\xiiptbfs
     \scriptscriptfont\bfsfam=\viiiptbf
     \@fontstyleinit
     \def\rm{\xviiptrm\fam=\z@}%
     \def\it{\xviiptit\fam=\itfam}%
     \def\bf{\xviiptbf\fam=\bffam}%
     \def\bfs{\xviiptbfs\fam=\bfsfam}%
     \def\oldstyle{\xviiptmit\fam=\@ne}%
     \rm\fi}


\font\xxiptrm=cmr17  scaled\magstep1


\def\xxipt{\ifmmode\err@badsizechange\else
     \@mathfontinit
     \@fontstyleinit
     \def\rm{\xxiptrm\fam=\z@}%
     \rm\fi}


\font\xxvptrm=cmr17  scaled\magstep2


\def\xxvpt{\ifmmode\err@badsizechange\else
     \@mathfontinit
     \@fontstyleinit
     \def\rm{\xxvptrm\fam=\z@}%
     \rm\fi}




\message{Loading jyTeX macros...}

\message{modifications to plain.tex,}


\def\newcount{\alloc@0\count\countdef\insc@unt}
\def\newdimen{\alloc@1\dimen\dimendef\insc@unt}
\def\newskip{\alloc@2\skip\skipdef\insc@unt}
\def\newmuskip{\alloc@3\muskip\muskipdef\@cclvi}
\def\newbox{\alloc@4\box\chardef\insc@unt}
\def\newtoks{\alloc@5\toks\toksdef\@cclvi}
\def\newhelp#1#2{\newtoks#1\global#1\expandafter{\csname#2\endcsname}}
\def\newread{\alloc@6\read\chardef\sixt@@n}
\def\newwrite{\alloc@7\write\chardef\sixt@@n}
\def\newfam{\alloc@8\fam\chardef\sixt@@n}
\def\newinsert#1{\global\advance\insc@unt by\m@ne
     \ch@ck0\insc@unt\count
     \ch@ck1\insc@unt\dimen
     \ch@ck2\insc@unt\skip
     \ch@ck4\insc@unt\box
     \allocationnumber=\insc@unt
     \global\chardef#1=\allocationnumber
     \wlog{\string#1=\string\insert\the\allocationnumber}}
\def\newif#1{\count@\escapechar \escapechar\m@ne
     \expandafter\expandafter\expandafter
          \xdef\@if#1{true}{\let\noexpand#1=\noexpand\iftrue}%
     \expandafter\expandafter\expandafter
          \xdef\@if#1{false}{\let\noexpand#1=\noexpand\iffalse}%
     \global\@if#1{false}\escapechar=\count@}


\newlinechar=`\^^J
\overfullrule=0pt




\let\itfam=\undefined

\let\bffam=\undefined

\count18=3


\chardef\sharps="19


\mathchardef\alpha="710B \mathchardef\beta="710C \mathchardef\gamma="710D
\mathchardef\delta="710E \mathchardef\epsilon="710F \mathchardef\zeta="7110
\mathchardef\eta="7111 \mathchardef\theta="7112 \mathchardef\iota="7113
\mathchardef\kappa="7114 \mathchardef\lambda="7115 \mathchardef\mu="7116
\mathchardef\nu="7117 \mathchardef\xi="7118 \mathchardef\pi="7119
\mathchardef\rho="711A \mathchardef\sigma="711B \mathchardef\tau="711C
\mathchardef\upsilon="711D \mathchardef\phi="711E \mathchardef\chi="711F
\mathchardef\psi="7120 \mathchardef\omega="7121 \mathchardef\varepsilon="7122
\mathchardef\vartheta="7123 \mathchardef\varpi="7124 \mathchardef\varrho="7125
\mathchardef\varsigma="7126 \mathchardef\varphi="7127 \mathchardef\imath="717B
\mathchardef\jmath="717C \mathchardef\ell="7160 \mathchardef\wp="717D
\mathchardef\partial="7140 \mathchardef\flat="715B \mathchardef\natural="715C
\mathchardef\sharp="715D



\def\angle{{\vbox{\ialign{$\m@th\scriptstyle##$\crcr
     \not\mathrel{\mkern14mu}\crcr
     \noalign{\nointerlineskip}
     \mkern2.5mu\leaders\hrule height.34\rp@\hfill\mkern2.5mu\crcr}}}}
\def\vdots{\vbox{\baselineskip4\rp@ \lineskiplimit\z@
     \kern6\rp@\hbox{.}\hbox{.}\hbox{.}}}
\def\ddots{\mathinner{\mkern1mu\raise7\rp@\vbox{\kern7\rp@\hbox{.}}\mkern2mu
     \raise4\rp@\hbox{.}\mkern2mu\raise\rp@\hbox{.}\mkern1mu}}
\def\overrightarrow#1{\vbox{\ialign{##\crcr
     \rightarrowfill\crcr
     \noalign{\kern-\rp@\nointerlineskip}
     $\hfil\displaystyle{#1}\hfil$\crcr}}}
\def\overleftarrow#1{\vbox{\ialign{##\crcr
     \leftarrowfill\crcr
     \noalign{\kern-\rp@\nointerlineskip}
     $\hfil\displaystyle{#1}\hfil$\crcr}}}
\def\overbrace#1{\mathop{\vbox{\ialign{##\crcr
     \noalign{\kern3\rp@}
     \downbracefill\crcr
     \noalign{\kern3\rp@\nointerlineskip}
     $\hfil\displaystyle{#1}\hfil$\crcr}}}\limits}
\def\underbrace#1{\mathop{\vtop{\ialign{##\crcr
     $\hfil\displaystyle{#1}\hfil$\crcr
     \noalign{\kern3\rp@\nointerlineskip}
     \upbracefill\crcr
     \noalign{\kern3\rp@}}}}\limits}
\def\big#1{{\hbox{$\left#1\vbox to8.5\rp@ {}\right.\n@space$}}}
\def\Big#1{{\hbox{$\left#1\vbox to11.5\rp@ {}\right.\n@space$}}}
\def\bigg#1{{\hbox{$\left#1\vbox to14.5\rp@ {}\right.\n@space$}}}
\def\Bigg#1{{\hbox{$\left#1\vbox to17.5\rp@ {}\right.\n@space$}}}
\def\@vereq#1#2{\lower.5\rp@\vbox{\baselineskip\z@skip\lineskip-.5\rp@
     \ialign{$\m@th#1\hfil##\hfil$\crcr#2\crcr=\crcr}}}
\def\rlh@#1{\vcenter{\hbox{\ooalign{\raise2\rp@
     \hbox{$#1\rightharpoonup$}\crcr
     $#1\leftharpoondown$}}}}
\def\bordermatrix#1{\begingroup\m@th
     \setbox\z@\vbox{%
          \def\cr{\crcr\noalign{\kern2\rp@\global\let\cr\endline}}%
          \ialign{$##$\hfil\kern2\rp@\kern\p@renwd
               &\thinspace\hfil$##$\hfil&&\quad\hfil$##$\hfil\crcr
               \omit\strut\hfil\crcr
               \noalign{\kern-\baselineskip}%
               #1\crcr\omit\strut\cr}}%
     \setbox\tw@\vbox{\unvcopy\z@\global\setbox\@ne\lastbox}%
     \setbox\tw@\hbox{\unhbox\@ne\unskip\global\setbox\@ne\lastbox}%
     \setbox\tw@\hbox{$\kern\wd\@ne\kern-\p@renwd\left(\kern-\wd\@ne
          \global\setbox\@ne\vbox{\box\@ne\kern2\rp@}%
          \vcenter{\kern-\ht\@ne\unvbox\z@\kern-\baselineskip}%
          \,\right)$}%
     \null\;\vbox{\kern\ht\@ne\box\tw@}\endgroup}
\def\endinsert{\egroup
     \if@mid\dimen@\ht\z@
          \advance\dimen@\dp\z@
          \advance\dimen@12\rp@
          \advance\dimen@\pagetotal
          \ifdim\dimen@>\pagegoal\@midfalse\p@gefalse\fi
     \fi
     \if@mid\bigskip\box\z@
          \bigbreak
     \else\insert\topins{\penalty100 \splittopskip\z@skip
               \splitmaxdepth\maxdimen\floatingpenalty\z@
               \ifp@ge\dimen@\dp\z@
                    \vbox to\vsize{\unvbox\z@\kern-\dimen@}%
               \else\box\z@\nobreak\bigskip
               \fi}%
     \fi
     \endgroup}


\def\cases#1{\left\{\,\vcenter{\m@th
     \ialign{$##\hfil$&\quad##\hfil\crcr#1\crcr}}\right.}
\def\matrix#1{\null\,\vcenter{\m@th
     \ialign{\hfil$##$\hfil&&\quad\hfil$##$\hfil\crcr
          \mathstrut\crcr
          \noalign{\kern-\baselineskip}
          #1\crcr
          \mathstrut\crcr
          \noalign{\kern-\baselineskip}}}\,}


\newif\ifraggedbottom

\def\raggedbottom{\ifraggedbottom\else
     \advance\topskip by\z@ plus60pt \raggedbottomtrue\fi}%
\def\normalbottom{\ifraggedbottom
     \advance\topskip by\z@ plus-60pt \raggedbottomfalse\fi}

\message{hacks,}


\toksdef\toks@i=1 \toksdef\toks@ii=2


\def\TeX{T\kern-.1667em \lower.5ex \hbox{E}\kern-.125em X\null}
\def\jyTeX{{\leavevmode
     \raise.587ex \hbox{\it\j}\kern-.1em \lower.048ex \hbox{\it y}\kern-.12em
     \TeX}}

\let\then=\iftrue
\def\ifnoarg#1\then{\def\hack@{#1}\ifx\hack@\empty}
\def\ifundefined#1\then{%
     \expandafter\ifx\csname\expandafter\blank\string#1\endcsname\relax}
\def\useif#1\then{\csname#1\endcsname}
\def\usename#1{\csname#1\endcsname}
\def\useafter#1#2{\expandafter#1\csname#2\endcsname}

\long\def\loop#1\repeat{\def\@iterate{#1\expandafter\@iterate\fi}\@iterate
     \let\@iterate=\relax}

\let\TeXend=\end
\def\begin#1{\begingroup\def\@@blockname{#1}\usename{begin#1}}
\def\end#1{\usename{end#1}\def\hack@{#1}%
     \ifx\@@blockname\hack@
          \endgroup
     \else\err@badgroup\hack@\@@blockname
     \fi}
\def\@@blockname{}

\def\defaultoption[#1]#2{%
     \def\hack@{\ifx\hack@ii[\toks@={#2}\else\toks@={#2[#1]}\fi\the\toks@}%
     \futurelet\hack@ii\hack@}

\def\markup#1{\let\@@marksf=\empty
     \ifhmode\edef\@@marksf{\spacefactor=\the\spacefactor\relax}\/\fi
     ${}^{\hbox{\subscriptfonts#1}}$\@@marksf}


\newtoks\shortyear
\newtoks\militaryhour
\newtoks\standardhour
\newtoks\minute
\newtoks\amorpm

\def\settime{\count@=\time\divide\count@ by60
     \militaryhour=\expandafter{\number\count@}%
     {\multiply\count@ by-60 \advance\count@ by\time
          \xdef\hack@{\ifnum\count@<10 0\fi\number\count@}}%
     \minute=\expandafter{\hack@}%
     \ifnum\count@<12
          \amorpm={am}
     \else\amorpm={pm}
          \ifnum\count@>12 \advance\count@ by-12 \fi
     \fi
     \standardhour=\expandafter{\number\count@}%
     \def\hack@19##1##2{\shortyear={##1##2}}%
          \expandafter\hack@\the\year}

\def\monthword#1{%
     \ifcase#1
          $\bullet$\err@badcountervalue{monthword}%
          \or January\or February\or March\or April\or May\or June%
          \or July\or August\or September\or October\or November\or December%
     \else$\bullet$\err@badcountervalue{monthword}%
     \fi}

\def\monthabbr#1{%
     \ifcase#1
          $\bullet$\err@badcountervalue{monthabbr}%
          \or Jan\or Feb\or Mar\or Apr\or May\or Jun%
          \or Jul\or Aug\or Sep\or Oct\or Nov\or Dec%
     \else$\bullet$\err@badcountervalue{monthabbr}%
     \fi}

\def\militarytime{\the\militaryhour:\the\minute}
\def\standardtime{\the\standardhour:\the\minute}


\def\@setnumstyle#1#2{\expandafter\global\expandafter\expandafter
     \expandafter\let\expandafter\expandafter
     \csname @\expandafter\blank\string#1style\endcsname
     \csname#2\endcsname}
\def\numstyle#1{\usename{@\expandafter\blank\string#1style}#1}
\def\ifblank#1\then{\useafter\ifx{@\expandafter\blank\string#1}\blank}

\def\blank#1{}

\def\Roman#1{\expandafter\uppercase\expandafter{\romannumeral#1}}
\def\alphabetic#1{%
     \ifcase#1
          $\bullet$\err@badcountervalue{alphabetic}%
          \or a\or b\or c\or d\or e\or f\or g\or h\or i\or j\or k\or l\or m%
          \or n\or o\or p\or q\or r\or s\or t\or u\or v\or w\or x\or y\or z%
     \else$\bullet$\err@badcountervalue{alphabetic}%
     \fi}
\def\Alphabetic#1{\expandafter\uppercase\expandafter{\alphabetic{#1}}}
\def\symbols#1{%
     \ifcase#1
          $\bullet$\err@badcountervalue{symbols}%
          \or*\or\dag\or\ddag\or\S\or$\|$%
          \or**\or\dag\dag\or\ddag\ddag\or\S\S\or$\|\|$%
     \else$\bullet$\err@badcountervalue{symbols}%
     \fi}


\catcode`\^^?=13 \def^^?{\relax}

\def\trimleading#1\to#2{\edef#2{#1}%
     \expandafter\@trimleading\expandafter#2#2^^?^^?}
\def\@trimleading#1#2#3^^?{\ifx#2^^?\def#1{}\else\def#1{#2#3}\fi}

\def\trimtrailing#1\to#2{\edef#2{#1}%
     \expandafter\@trimtrailing\expandafter#2#2^^? ^^?\relax}
\def\@trimtrailing#1#2 ^^?#3{\ifx#3\relax\toks@={}%
     \else\def#1{#2}\toks@={\trimtrailing#1\to#1}\fi
     \the\toks@}

\def\trim#1\to#2{\trimleading#1\to#2\trimtrailing#2\to#2}

\catcode`\^^?=15


\long\def\additemL#1\to#2{\toks@={\^^\{#1}}\toks@ii=\expandafter{#2}%
     \xdef#2{\the\toks@\the\toks@ii}}

\long\def\additemR#1\to#2{\toks@={\^^\{#1}}\toks@ii=\expandafter{#2}%
     \xdef#2{\the\toks@ii\the\toks@}}

\def\getitemL#1\to#2{\expandafter\@getitemL#1\hack@#1#2}
\def\@getitemL\^^\#1#2\hack@#3#4{\def#4{#1}\def#3{#2}}

\message{font macros,}


\newdimen\rp@
\newcount\@@sizeindex \@@sizeindex=0
\newcount\@@factori
\newcount\@@factorii
\newcount\@@factoriii
\newcount\@@factoriv

\countdef\maxfam=18
\newfam\itfam
\newfam\bffam
\newfam\bfsfam
\newfam\bmitfam

\def\@mathfontinit{\count@=4
     \loop\textfont\count@=\nullfont
          \scriptfont\count@=\nullfont
          \scriptscriptfont\count@=\nullfont
          \ifnum\count@<\maxfam\advance\count@ by\@ne
     \repeat}

\def\@fontstyleinit{%
     \def\it{\err@fontnotavailable\it}%
     \def\bf{\err@fontnotavailable\bf}%
     \def\bfs{\err@bfstobf}%
     \def\bmit{\err@fontnotavailable\bmit}%
     \def\sc{\err@fontnotavailable\sc}%
     \def\sl{\err@sltoit}%
     \def\ss{\err@fontnotavailable\ss}%
     \def\tt{\err@fontnotavailable\tt}}

\def\@parameterinit#1{\rm\rp@=.1em \@getscaling{#1}%
     \let\^^\=\@doscaling\scalingskipslist
     \setbox\strutbox=\hbox{\vrule
          height.708\baselineskip depth.292\baselineskip width\z@}}

\def\@getfactor#1#2#3#4{\@@factori=#1 \@@factorii=#2
     \@@factoriii=#3 \@@factoriv=#4}

\def\@getscaling#1{\count@=#1 \advance\count@ by-\@@sizeindex\@@sizeindex=#1
     \ifnum\count@<0
          \let\@mulordiv=\divide
          \let\@divormul=\multiply
          \multiply\count@ by\m@ne
     \else\let\@mulordiv=\multiply
          \let\@divormul=\divide
     \fi
     \edef\@@scratcha{\ifcase\count@                {1}{1}{1}{1}\or
          {1}{7}{23}{3}\or     {2}{5}{3}{1}\or      {9}{89}{13}{1}\or
          {6}{25}{6}{1}\or     {8}{71}{14}{1}\or    {6}{25}{36}{5}\or
          {1}{7}{53}{4}\or     {12}{125}{108}{5}\or {3}{14}{53}{5}\or
          {6}{41}{17}{1}\or    {13}{31}{13}{2}\or   {9}{107}{71}{2}\or
          {11}{139}{124}{3}\or {1}{6}{43}{2}\or     {10}{107}{42}{1}\or
          {1}{5}{43}{2}\or     {5}{69}{65}{1}\or    {11}{97}{91}{2}\fi}%
     \expandafter\@getfactor\@@scratcha}

\def\@doscaling#1{\@mulordiv#1by\@@factori\@divormul#1by\@@factorii
     \@mulordiv#1by\@@factoriii\@divormul#1by\@@factoriv}


\newskip\headskip
\newskip\footskip

\def\typesize=#1pt{\count@=#1 \advance\count@ by-10
     \ifcase\count@
          \@setsizex\or\err@badtypesize\or
          \@setsizexii\or\err@badtypesize\or
          \@setsizexiv
     \else\err@badtypesize
     \fi}

\def\@setsizex{\getixpt
     \def\subsubscriptfonts{\vpt}%
          \def\subsubscriptsize{\vpt\@parameterinit{-8}}%
     \def\subscriptfonts{\viipt}\def\subscriptsize{\viipt\@parameterinit{-4}}%
     \def\footnotefonts{\viiipt}\def\footnotesize{\viiipt\@parameterinit{-2}}%
      \def\smallfonts{\ixpt}\def\smallsize{\ixpt\@parameterinit{-1}}%
     \def\normalfonts{\xpt}\def\normalsize{\xpt\@parameterinit{0}}%
     \def\bigfonts{\xiipt}\def\bigsize{\xiipt\@parameterinit{2}}%
     \def\Bigfonts{\xivpt}\def\Bigsize{\xivpt\@parameterinit{4}}%
     \def\biggfonts{\xviipt}\def\biggsize{\xviipt\@parameterinit{6}}%
     \def\Biggfonts{\xxipt}\def\Biggsize{\xxipt\@parameterinit{8}}%
     \def\tinyfonts{\vpt}\def\tinysize{\vpt\@parameterinit{-8}}%
     \def\HUGEFONTS{\xxvpt}\def\HUGESIZE{\xxvpt\@parameterinit{10}}%
     \normalsize\fixedskipslist}

\def\@setsizexii{\getxipt
     \def\subsubscriptfonts{\vipt}%
          \def\subsubscriptsize{\vipt\@parameterinit{-6}}%
     \def\subscriptfonts{\viiipt}%
          \def\subscriptsize{\viiipt\@parameterinit{-2}}%
     \def\footnotefonts{\xpt}\def\footnotesize{\xpt\@parameterinit{0}}%
     \def\smallfonts{\xipt}\def\smallsize{\xipt\@parameterinit{1}}%
     \def\normalfonts{\xiipt}\def\normalsize{\xiipt\@parameterinit{2}}%
     \def\bigfonts{\xivpt}\def\bigsize{\xivpt\@parameterinit{4}}%
     \def\Bigfonts{\xviipt}\def\Bigsize{\xviipt\@parameterinit{6}}%
     \def\biggfonts{\xxipt}\def\biggsize{\xxipt\@parameterinit{8}}%
     \def\Biggfonts{\xxvpt}\def\Biggsize{\xxvpt\@parameterinit{10}}%
     \def\tinyfonts{\vpt}\def\tinysize{\vpt\@parameterinit{-8}}%
     \def\HUGEFONTS{\xxvpt}\def\HUGESIZE{\xxvpt\@parameterinit{10}}%
     \normalsize\fixedskipslist}

\def\@setsizexiv{\getxiiipt
     \def\subsubscriptfonts{\viipt}%
          \def\subsubscriptsize{\viipt\@parameterinit{-4}}%
     \def\subscriptfonts{\xpt}\def\subscriptsize{\xpt\@parameterinit{0}}%
     \def\footnotefonts{\xiipt}\def\footnotesize{\xiipt\@parameterinit{2}}%
     \def\smallfonts{\xiiipt}\def\smallsize{\xiiipt\@parameterinit{3}}%
     \def\normalfonts{\xivpt}\def\normalsize{\xivpt\@parameterinit{4}}%
     \def\bigfonts{\xviipt}\def\bigsize{\xviipt\@parameterinit{6}}%
     \def\Bigfonts{\xxipt}\def\Bigsize{\xxipt\@parameterinit{8}}%
     \def\biggfonts{\xxvpt}\def\biggsize{\xxvpt\@parameterinit{10}}%
     \def\Biggfonts{\err@sizetoolarge\Biggfonts\HUGEFONTS}%
          \def\Biggsize{\err@sizetoolarge\Biggsize\HUGESIZE}%
     \def\tinyfonts{\vpt}\def\tinysize{\vpt\@parameterinit{-8}}%
     \def\HUGEFONTS{\xxvpt}\def\HUGESIZE{\xxvpt\@parameterinit{10}}%
     \normalsize\fixedskipslist}

\def\subsubscriptfonts{\vpt} \def\subsubscriptsize{\vpt\@parameterinit{-8}}
\def\subscriptfonts{\viipt}  \def\subscriptsize{\viipt\@parameterinit{-4}}
\def\footnotefonts{\viiipt}  \def\footnotesize{\viiipt\@parameterinit{-2}}
\def\smallfonts{\err@sizenotavailable\smallfonts}
                             \def\smallsize{\ixpt\@parameterinit{-1}}
\def\normalfonts{\xpt}       \def\normalsize{\xpt\@parameterinit{0}}
\def\bigfonts{\xiipt}        \def\bigsize{\xiipt\@parameterinit{2}}
\def\Bigfonts{\xivpt}        \def\Bigsize{\xivpt\@parameterinit{4}}
\def\biggfonts{\xviipt}      \def\biggsize{\xviipt\@parameterinit{6}}
\def\Biggfonts{\xxipt}       \def\Biggsize{\xxipt\@parameterinit{8}}
\def\tinyfonts{\vpt}         \def\tinysize{\vpt\@parameterinit{-8}}
\def\HUGEFONTS{\xxvpt}       \def\HUGESIZE{\xxvpt\@parameterinit{10}}

\message{document layout,}


\newtoks\everyoutput \everyoutput={}
\newdimen\depthofpage
\newcount\pagenum \pagenum=0

\newdimen\oddtopmargin  \newdimen\eventopmargin
\newdimen\oddleftmargin \newdimen\evenleftmargin
\newtoks\oddhead        \newtoks\evenhead
\newtoks\oddfoot        \newtoks\evenfoot

\def\topmargin{\afterassignment\@seteventop\oddtopmargin}
\def\leftmargin{\afterassignment\@setevenleft\oddleftmargin}
\def\head{\afterassignment\@setevenhead\oddhead}
\def\foot{\afterassignment\@setevenfoot\oddfoot}

\def\@seteventop{\eventopmargin=\oddtopmargin}
\def\@setevenleft{\evenleftmargin=\oddleftmargin}
\def\@setevenhead{\evenhead=\oddhead}
\def\@setevenfoot{\evenfoot=\oddfoot}

\def\pagenumstyle#1{\@setnumstyle\pagenum{#1}}

\newif\ifdraft
\def\draft{\drafttrue\leftmargin=.5in \overfullrule=5pt }

\def\outputstyle#1{\global\expandafter\let\expandafter
          \@outputstyle\csname#1output\endcsname
     \usename{#1setup}}

\output={\@outputstyle}

\def\normaloutput{\the\everyoutput
     \global\advance\pagenum by\@ne
     \ifodd\pagenum
          \voffset=\oddtopmargin \hoffset=\oddleftmargin
     \else\voffset=\eventopmargin \hoffset=\evenleftmargin
     \fi
     \advance\voffset by-1in  \advance\hoffset by-1in
     \count0=\pagenum
     \expandafter\shipout\pagebox
     \ifnum\outputpenalty>-\@MM\else\dosupereject\fi}

\newdimen\fullhsize
\newbox\leftpage
\newcount\leftpagenum
\newcount\outputpagenum \outputpagenum=0
\let\leftorright=L

\def\twoupoutput{\the\everyoutput
     \global\advance\pagenum by\@ne
     \if L\leftorright
          \global\setbox\leftpage=\leftline{\pagebox}%
          \global\leftpagenum=\pagenum
          \global\let\leftorright=R%
     \else\global\advance\outputpagenum by\@ne
          \ifodd\outputpagenum
               \voffset=\oddtopmargin \hoffset=\oddleftmargin
          \else\voffset=\eventopmargin \hoffset=\evenleftmargin
          \fi
          \advance\voffset by-1in  \advance\hoffset by-1in
          \count0=\leftpagenum \count1=\pagenum
          \shipout\vbox{\hbox to\fullhsize
               {\box\leftpage\hfil\leftline{\pagebox}}}%
          \global\let\leftorright=L%
     \fi
     \ifnum\outputpenalty>-\@MM
     \else\dosupereject
          \if R\leftorright
               \globaldefs=\@ne\head={\hfil}\foot={\hfil}\globaldefs=\z@
               \null\newpage
          \fi
     \fi}

\def\pagebox{\vbox{\makeheadline\pagebody\makefootline}}

\def\makeheadline{%
     \vbox to\z@{\baselinestretch=\@m
          \vskip\topskip\vskip-.708\baselineskip\vskip-\headskip
          \line{\vbox to\ht\strutbox{}%
               \ifodd\pagenum\the\oddhead\else\the\evenhead\fi}%
          \vss}%
     \nointerlineskip}

\def\pagebody{\vbox to\vsize{%
     \boxmaxdepth\maxdepth
     \ifvoid\topins\else\unvbox\topins\fi
     \depthofpage=\dp255
     \unvbox255
     \ifraggedbottom\kern-\depthofpage\vfil\fi
     \ifvoid\footins
     \else\vskip\skip\footins
          \footnoterule
          \unvbox\footins
          \vskip-\footnoteskip
     \fi}}

\def\makefootline{\baselineskip=\footskip
     \line{\ifodd\pagenum\the\oddfoot\else\the\evenfoot\fi}}


\newskip\abovechapterskip
\newskip\belowchapterskip
\newskip\abovesectionskip
\newskip\belowsectionskip
\newskip\abovesubsectionskip
\newskip\belowsubsectionskip

\def\chapterstyle#1{\global\expandafter\let\expandafter\@chapterstyle
     \csname#1text\endcsname}
\def\sectionstyle#1{\global\expandafter\let\expandafter\@sectionstyle
     \csname#1text\endcsname}
\def\subsectionstyle#1{\global\expandafter\let\expandafter\@subsectionstyle
     \csname#1text\endcsname}

\def\chapter#1{%
     \ifdim\lastskip=17sp \else\chapterbreak\vskip\abovechapterskip\fi
     \@chapterstyle{\ifblank\chapternumstyle\then
          \else\newchapternum=\next\chapternumformat\ \fi#1}%
     \nobreak\vskip\belowchapterskip\vskip17sp }

\def\section#1{%
     \ifdim\lastskip=17sp \else\sectionbreak\vskip\abovesectionskip\fi
     \@sectionstyle{\ifblank\sectionnumstyle\then
          \else\newsectionnum=\next\sectionnumformat\ \fi#1}%
     \nobreak\vskip\belowsectionskip\vskip17sp }

\def\subsection#1{%
     \ifdim\lastskip=17sp \else\subsectionbreak\vskip\abovesubsectionskip\fi
     \@subsectionstyle{\ifblank\subsectionnumstyle\then
          \else\newsubsectionnum=\next\subsectionnumformat\ \fi#1}%
     \nobreak\vskip\belowsubsectionskip\vskip17sp }


\let\TeXunderline=\underline
\let\TeXoverline=\overline
\def\underline#1{\relax\ifmmode\TeXunderline{#1}\else
     $\TeXunderline{\hbox{#1}}$\fi}
\def\overline#1{\relax\ifmmode\TeXoverline{#1}\else
     $\TeXoverline{\hbox{#1}}$\fi}

\def\baselinestretch{\afterassignment\@baselinestretch\count@}
\def\@baselinestretch{\baselineskip=\normalbaselineskip
     \divide\baselineskip by\@m\baselineskip=\count@\baselineskip
     \setbox\strutbox=\hbox{\vrule
          height.708\baselineskip depth.292\baselineskip width\z@}%
     \bigskipamount=\the\baselineskip
          plus.25\baselineskip minus.25\baselineskip
     \medskipamount=.5\baselineskip
          plus.125\baselineskip minus.125\baselineskip
     \smallskipamount=.25\baselineskip
          plus.0625\baselineskip minus.0625\baselineskip}

\def\\{\ifhmode\ifnum\lastpenalty=-\@M\else\hfil\penalty-\@M\fi\fi
     \ignorespaces}
\def\newpage{\vfil\break}

\def\lefttext#1{\par{\@text\leftskip=\z@\rightskip=\centering
     \noindent#1\par}}
\def\righttext#1{\par{\@text\leftskip=\centering\rightskip=\z@
     \noindent#1\par}}
\def\centertext#1{\par{\@text\leftskip=\centering\rightskip=\centering
     \noindent#1\par}}
\def\@text{\parindent=\z@ \parfillskip=\z@ \everypar={}%
     \spaceskip=.3333em \xspaceskip=.5em
     \def\\{\ifhmode\ifnum\lastpenalty=-\@M\else\penalty-\@M\fi\fi
          \ignorespaces}}

\def\beginleft{\par\@text\leftskip=\z@ \rightskip=\centering}
     
\def\beginright{\par\@text\leftskip=\centering\rightskip=\z@ }
     
\def\begincenter{\par\@text\leftskip=\centering\rightskip=\centering}

\def\beginnarrow{\defaultoption[\parindent]\@beginnarrow}
\def\@beginnarrow[#1]{\par\advance\leftskip by#1\advance\rightskip by#1}

\begingroup
\catcode`\[=1 \catcode`\{=11 \gdef\beginignore[\endgroup\bgroup
     \catcode`\e=0 \catcode`\\=12 \catcode`\{=11 \catcode`\f=12 \let\or=\relax
     \let\nd{ignor=\fi \let\}=\egroup
     \iffalse}
\endgroup

\long\def\marginnote#1{\leavevmode
     \edef\@marginsf{\spacefactor=\the\spacefactor\relax}%
     \ifdraft\strut\vadjust{%
          \hbox to\z@{\hskip\hsize\hskip.1in
               \vbox to\z@{\vskip-\dp\strutbox
                    \marginnoteformat
                    \vskip-\ht\strutbox
                    \noindent\strut#1\par
                    \vss}%
               \hss}}%
     \fi
     \@marginsf}


\newtoks\everybye \everybye={\par\vfil}
\outer\def\bye{\the\everybye
     \footnotecheck
     \prelabelcheck
     \streamcheck
     \supereject
     \TeXend}

\message{footnotes,}

\newcount\footnotenum \footnotenum=0
\newskip\footnoteskip
\let\@footnotelist=\empty

\def\footnotenumstyle#1{\@setnumstyle\footnotenum{#1}%
     \useafter\ifx{@footnotenumstyle}\symbols
          \global\let\@footup=\empty
     \else\global\let\@footup=\markup
     \fi}

\def\footnote{\footnotecheck\defaultoption[]\@footnote}
\def\@footnote[#1]{\@footnotemark[#1]\@footnotetext}

\def\footnotemark{\defaultoption[]\@footnotemark}
\def\@footnotemark[#1]{\let\@footsf=\empty
     \ifhmode\edef\@footsf{\spacefactor=\the\spacefactor\relax}\/\fi
     \ifnoarg#1\then
          \global\advance\footnotenum by\@ne
          \@footup{\footnotenumformat}%
          \edef\@@foota{\footnotenum=\the\footnotenum\relax}%
          \expandafter\additemR\expandafter\@footup\expandafter
               {\@@foota\footnotenumformat}\to\@footnotelist
          \global\let\@footnotelist=\@footnotelist
     \else\markup{#1}%
          \additemR\markup{#1}\to\@footnotelist
          \global\let\@footnotelist=\@footnotelist
     \fi
     \@footsf}

\def\footnotetext{%
     \ifx\@footnotelist\empty\err@extrafootnotetext\else\@footnotetext\fi}
\def\@footnotetext{%
     \getitemL\@footnotelist\to\@@foota
     \global\let\@footnotelist=\@footnotelist
     \insert\footins\bgroup
     \footnoteformat
     \splittopskip=\ht\strutbox\splitmaxdepth=\dp\strutbox
     \interlinepenalty=\interfootnotelinepenalty\floatingpenalty=\@MM
     \noindent\llap{\@@foota}\strut
     \bgroup\aftergroup\@footnoteend
     \let\@@scratcha=}
\def\@footnoteend{\strut\par\vskip\footnoteskip\egroup}

\def\footnoterule{\normalfonts
     \kern-.3em \hrule width2in height.04em \kern .26em }

\def\footnotecheck{%
     \ifx\@footnotelist\empty
     \else\err@extrafootnotemark
          \global\let\@footnotelist=\empty
     \fi}

\message{labels,}

\let\@@labeldef=\xdef
\newif\if@labelfile
\newwrite\@labelfile
\let\@prelabellist=\empty

\def\label#1#2{\trim#1\to\@@labarg\edef\@@labtext{#2}%
     \edef\@@labname{lab@\@@labarg}%
     \useafter\ifundefined\@@labname\then\else\@yeslab\fi
     \useafter\@@labeldef\@@labname{#2}%
     \ifstreaming
          \expandafter\toks@\expandafter\expandafter\expandafter
               {\csname\@@labname\endcsname}%
          \immediate\write\streamout{\noexpand\label{\@@labarg}{\the\toks@}}%
     \fi}
\def\@yeslab{%
     \useafter\ifundefined{if\@@labname}\then
          \err@labelredef\@@labarg
     \else\useif{if\@@labname}\then
               \err@labelredef\@@labarg
          \else\global\usename{\@@labname true}%
               \useafter\ifundefined{pre\@@labname}\then
               \else\useafter\ifx{pre\@@labname}\@@labtext
                    \else\err@badlabelmatch\@@labarg
                    \fi
               \fi
               \if@labelfile
               \else\global\@labelfiletrue
                    \immediate\write\sixt@@n{--> Creating file \jobname.lab}%
                    \immediate\openout\@labelfile=\jobname.lab
               \fi
               \immediate\write\@labelfile
                    {\noexpand\prelabel{\@@labarg}{\@@labtext}}%
          \fi
     \fi}

\def\putlab#1{\trim#1\to\@@labarg\edef\@@labname{lab@\@@labarg}%
     \useafter\ifundefined\@@labname\then\@nolab\else\usename\@@labname\fi}
\def\@nolab{%
     \useafter\ifundefined{pre\@@labname}\then
          \undefinedlabelformat
          \err@needlabel\@@labarg
          \useafter\xdef\@@labname{\undefinedlabelformat}%
     \else\usename{pre\@@labname}%
          \useafter\xdef\@@labname{\usename{pre\@@labname}}%
     \fi
     \useafter\newif{if\@@labname}%
     \expandafter\additemR\@@labarg\to\@prelabellist}

\def\prelabel#1{\useafter\gdef{prelab@#1}}

\def\ifundefinedlabel#1\then{%
     \expandafter\ifx\csname lab@#1\endcsname\relax}
\def\useiflab#1\then{\csname iflab@#1\endcsname}

\def\prelabelcheck{{%
     \def\^^\##1{\useiflab{##1}\then\else\err@undefinedlabel{##1}\fi}%
     \@prelabellist}}

\message{equation numbering,}

\newcount\chapternum
\newcount\sectionnum
\newcount\subsectionnum
\newcount\equationnum
\newcount\subequationnum
\newcount\figurenum
\newcount\subfigurenum
\newcount\tablenum
\newcount\subtablenum

\newif\if@subeqncount
\newif\if@subfigcount
\newif\if@subtblcount

\def\newchapternum{\newsectionnum=\z@\@resetnum\chapternum}
\def\newsectionnum{\newsubsectionnum=\z@\@resetnum\sectionnum}
\def\newsubsectionnum{\newequationnum=\z@\newfigurenum=\z@\newtablenum=\z@
     \@resetnum\subsectionnum}
\def\newequationnum{\newsubequationnum=\z@\@resetnum\equationnum}
\def\newsubequationnum{\@resetnum\subequationnum}
\def\newfigurenum{\newsubfigurenum=\z@\@resetnum\figurenum}
\def\newsubfigurenum{\@resetnum\subfigurenum}
\def\newtablenum{\newsubtablenum=\z@\@resetnum\tablenum}
\def\newsubtablenum{\@resetnum\subtablenum}

\def\@resetnum#1{\global\advance#1by1 \edef\next{\the#1\relax}\global#1}

\newchapternum=0

\def\chapternumstyle#1{\@setnumstyle\chapternum{#1}}
\def\sectionnumstyle#1{\@setnumstyle\sectionnum{#1}}
\def\subsectionnumstyle#1{\@setnumstyle\subsectionnum{#1}}
\def\equationnumstyle#1{\@setnumstyle\equationnum{#1}}
\def\subequationnumstyle#1{\@setnumstyle\subequationnum{#1}%
     \ifblank\subequationnumstyle\then\global\@subeqncountfalse\fi
     \ignorespaces}
\def\figurenumstyle#1{\@setnumstyle\figurenum{#1}}
\def\subfigurenumstyle#1{\@setnumstyle\subfigurenum{#1}%
     \ifblank\subfigurenumstyle\then\global\@subfigcountfalse\fi
     \ignorespaces}
\def\tablenumstyle#1{\@setnumstyle\tablenum{#1}}
\def\subtablenumstyle#1{\@setnumstyle\subtablenum{#1}%
     \ifblank\subtablenumstyle\then\global\@subtblcountfalse\fi
     \ignorespaces}

\def\eqnlabel#1{%
     \if@subeqncount
          \newsubequationnum=\next
     \else\newequationnum=\next
          \ifblank\subequationnumstyle\then
          \else\global\@subeqncounttrue
               \newsubequationnum=\@ne
          \fi
     \fi
     \label{#1}{\puteqnformat}(\puteqn{#1})%
     \ifdraft\rlap{\hskip.1in{\tt#1}}\fi}

\let\puteqn=\putlab

\def\equation#1#2{\useafter\gdef{eqn@#1}{#2\eqno\eqnlabel{#1}}}
\def\Equation#1{\useafter\gdef{eqn@#1}}

\def\putequation#1{\useafter\ifundefined{eqn@#1}\then
     \err@undefinedeqn{#1}\else\usename{eqn@#1}\fi}

\def\eqnseriesstyle#1{\gdef\@eqnseriesstyle{#1}}
\def\begineqnseries{\subequationnumstyle{\@eqnseriesstyle}%
     \defaultoption[]\@begineqnseries}
\def\@begineqnseries[#1]{\edef\@@eqnname{#1}}
\def\endeqnseries{\subequationnumstyle{blank}%
     \expandafter\ifnoarg\@@eqnname\then
     \else\label\@@eqnname{\puteqnformat}%
     \fi
     \aftergroup\ignorespaces}

\def\figlabel#1{%
     \if@subfigcount
          \newsubfigurenum=\next
     \else\newfigurenum=\next
          \ifblank\subfigurenumstyle\then
          \else\global\@subfigcounttrue
               \newsubfigurenum=\@ne
          \fi
     \fi
     \label{#1}{\putfigformat}\putfig{#1}%
     {\def\marginnoteformat{\tt}\marginnote{#1}}}

\let\putfig=\putlab

\def\figseriesstyle#1{\gdef\@figseriesstyle{#1}}
\def\beginfigseries{\subfigurenumstyle{\@figseriesstyle}%
     \defaultoption[]\@beginfigseries}
\def\@beginfigseries[#1]{\edef\@@figname{#1}}
\def\endfigseries{\subfigurenumstyle{blank}%
     \expandafter\ifnoarg\@@figname\then
     \else\label\@@figname{\putfigformat}%
     \fi
     \aftergroup\ignorespaces}

\def\tbllabel#1{%
     \if@subtblcount
          \newsubtablenum=\next
     \else\newtablenum=\next
          \ifblank\subtablenumstyle\then
          \else\global\@subtblcounttrue
               \newsubtablenum=\@ne
          \fi
     \fi
     \label{#1}{\puttblformat}\puttbl{#1}%
     {\def\marginnoteformat{\tt}\marginnote{#1}}}

\let\puttbl=\putlab

\def\tblseriesstyle#1{\gdef\@tblseriesstyle{#1}}
\def\begintblseries{\subtablenumstyle{\@tblseriesstyle}%
     \defaultoption[]\@begintblseries}
\def\@begintblseries[#1]{\edef\@@tblname{#1}}
\def\endtblseries{\subtablenumstyle{blank}%
     \expandafter\ifnoarg\@@tblname\then
     \else\label\@@tblname{\puttblformat}%
     \fi
     \aftergroup\ignorespaces}

\message{reference numbering,}

\newcount\referencenum \referencenum=0
\newcount\@@prerefcount \@@prerefcount=0
\newcount\@@thisref
\newcount\@@lastref
\newcount\@@loopref
\newcount\@@refseq
\newdimen\refnumindent
\let\@undefreflist=\empty

\def\referencenumstyle#1{\@setnumstyle\referencenum{#1}}

\def\referencestyle#1{\usename{@ref#1}}

\def\@refsequential{%
     \gdef\@refpredef##1{\global\advance\referencenum by\@ne
          \let\^^\=0\label{##1}{\^^\{\the\referencenum}}%
          \useafter\gdef{ref@\the\referencenum}{{##1}{\undefinedlabelformat}}}%
     \gdef\@reference##1##2{%
          \ifundefinedlabel##1\then
          \else\def\^^\####1{\global\@@thisref=####1\relax}\putlab{##1}%
               \useafter\gdef{ref@\the\@@thisref}{{##1}{##2}}%
          \fi}%
     \gdef\endputreferences{%
          \loop\ifnum\@@loopref<\referencenum
                    \advance\@@loopref by\@ne
                    \expandafter\expandafter\expandafter\@printreference
                         \csname ref@\the\@@loopref\endcsname
          \repeat
          \par}}

\def\@refpreordered{%
     \gdef\@refpredef##1{\global\advance\referencenum by\@ne
          \additemR##1\to\@undefreflist}%
     \gdef\@reference##1##2{%
          \ifundefinedlabel##1\then
          \else\global\advance\@@loopref by\@ne
               {\let\^^\=0\label{##1}{\^^\{\the\@@loopref}}}%
               \@printreference{##1}{##2}%
          \fi}
     \gdef\endputreferences{%
          \def\^^\####1{\useiflab{####1}\then
               \else\reference{####1}{\undefinedlabelformat}\fi}%
          \@undefreflist
          \par}}

\def\beginprereferences{\par
     \def\reference##1##2{\global\advance\referencenum by1\@ne
          \let\^^\=0\label{##1}{\^^\{\the\referencenum}}%
          \useafter\gdef{ref@\the\referencenum}{{##1}{##2}}}}
\def\endprereferences{\global\@@prerefcount=\the\referencenum\par}

\def\beginputreferences{\par
     \refnumindent=\z@\@@loopref=\z@
     \loop\ifnum\@@loopref<\referencenum
               \advance\@@loopref by\@ne
               \setbox\z@=\hbox{\referencenum=\@@loopref
                    \referencenumformat\enskip}%
               \ifdim\wd\z@>\refnumindent\refnumindent=\wd\z@\fi
     \repeat
     \putreferenceformat
     \@@loopref=\z@
     \loop\ifnum\@@loopref<\@@prerefcount
               \advance\@@loopref by\@ne
               \expandafter\expandafter\expandafter\@printreference
                    \csname ref@\the\@@loopref\endcsname
     \repeat
     \let\reference=\@reference}

\def\@printreference#1#2{\ifx#2\undefinedlabelformat\err@undefinedref{#1}\fi
     \noindent\ifdraft\rlap{\hskip\hsize\hskip.1in \tt#1}\fi
     \llap{\referencenum=\@@loopref\referencenumformat\enskip}#2\par}

\def\reference#1#2{{\par\refnumindent=\z@\putreferenceformat\noindent#2\par}}

\def\putref#1{\trim#1\to\@@refarg
     \expandafter\ifnoarg\@@refarg\then
          \toks@={\relax}%
     \else\@@lastref=-\@m\def\@@refsep{}\def\@more{\@nextref}%
          \toks@={\@nextref#1,,}%
     \fi\the\toks@}
\def\@nextref#1,{\trim#1\to\@@refarg
     \expandafter\ifnoarg\@@refarg\then
          \let\@more=\relax
     \else\ifundefinedlabel\@@refarg\then
               \expandafter\@refpredef\expandafter{\@@refarg}%
          \fi
          \def\^^\##1{\global\@@thisref=##1\relax}%
          \global\@@thisref=\m@ne
          \setbox\z@=\hbox{\putlab\@@refarg}%
     \fi
     \advance\@@lastref by\@ne
     \ifnum\@@lastref=\@@thisref\advance\@@refseq by\@ne\else\@@refseq=\@ne\fi
     \ifnum\@@lastref<\z@
     \else\ifnum\@@refseq<\thr@@
               \@@refsep\def\@@refsep{,}%
               \ifnum\@@lastref>\z@
                    \advance\@@lastref by\m@ne
                    {\referencenum=\@@lastref\putrefformat}%
               \else\undefinedlabelformat
               \fi
          \else\def\@@refsep{--}%
          \fi
     \fi
     \@@lastref=\@@thisref
     \@more}

\message{streaming,}

\newif\ifstreaming

\def\streamto{\defaultoption[\jobname]\@streamto}
\def\@streamto[#1]{\global\streamingtrue
     \immediate\write\sixt@@n{--> Streaming to #1.str}%
     \newwrite\streamout\immediate\openout\streamout=#1.str }

\def\streamfrom{\defaultoption[\jobname]\@streamfrom}
\def\@streamfrom[#1]{\newread\streamin\openin\streamin=#1.str
     \ifeof\streamin
          \expandafter\err@nostream\expandafter{#1.str}%
     \else\immediate\write\sixt@@n{--> Streaming from #1.str}%
          \let\@@labeldef=\gdef
          \ifstreaming
               \edef\@elc{\endlinechar=\the\endlinechar}%
               \endlinechar=\m@ne
               \loop\read\streamin to\@@scratcha
                    \ifeof\streamin
                         \streamingfalse
                    \else\toks@=\expandafter{\@@scratcha}%
                         \immediate\write\streamout{\the\toks@}%
                    \fi
                    \ifstreaming
               \repeat
               \@elc
               \input #1.str
               \streamingtrue
          \else\input #1.str
          \fi
          \let\@@labeldef=\xdef
     \fi}

\def\streamcheck{\ifstreaming
     \immediate\write\streamout{\pagenum=\the\pagenum}%
     \immediate\write\streamout{\footnotenum=\the\footnotenum}%
     \immediate\write\streamout{\referencenum=\the\referencenum}%
     \immediate\write\streamout{\chapternum=\the\chapternum}%
     \immediate\write\streamout{\sectionnum=\the\sectionnum}%
     \immediate\write\streamout{\subsectionnum=\the\subsectionnum}%
     \immediate\write\streamout{\equationnum=\the\equationnum}%
     \immediate\write\streamout{\subequationnum=\the\subequationnum}%
     \immediate\write\streamout{\figurenum=\the\figurenum}%
     \immediate\write\streamout{\subfigurenum=\the\subfigurenum}%
     \immediate\write\streamout{\tablenum=\the\tablenum}%
     \immediate\write\streamout{\subtablenum=\the\subtablenum}%
     \immediate\closeout\streamout
     \fi}


\def\err@badtypesize{%
     \errhelp={The limited availability of certain fonts requires^^J%
          that the base type size be 10pt, 12pt, or 14pt.^^J}%
     \errmessage{--> Illegal base type size}}

\def\err@badsizechange{\immediate\write\sixt@@n
     {--> Size change not allowed in math mode, ignored}}

\def\err@sizetoolarge#1{\immediate\write\sixt@@n
     {--> \noexpand#1 too big, substituting HUGE}}

\def\err@sizenotavailable#1{\immediate\write\sixt@@n
     {--> Size not available, \noexpand#1 ignored}}

\def\err@fontnotavailable#1{\immediate\write\sixt@@n
     {--> Font not available, \noexpand#1 ignored}}

\def\err@sltoit{\immediate\write\sixt@@n
     {--> Style \noexpand\sl not available, substituting \noexpand\it}%
     \it}

\def\err@bfstobf{\immediate\write\sixt@@n
     {--> Style \noexpand\bfs not available, substituting \noexpand\bf}%
     \bf}

\def\err@badgroup#1#2{%
     \errhelp={The block you have just tried to close was not the one^^J%
          most recently opened.^^J}%
     \errmessage{--> \noexpand\end{#1} doesn't match \noexpand\begin{#2}}}

\def\err@badcountervalue#1{\immediate\write\sixt@@n
     {--> Counter (#1) out of bounds}}

\def\err@extrafootnotemark{\immediate\write\sixt@@n
     {--> \noexpand\footnotemark command
          has no corresponding \noexpand\footnotetext}}

\def\err@extrafootnotetext{%
     \errhelp{You have given a \noexpand\footnotetext command without first
          specifying^^Ja \noexpand\footnotemark.^^J}%
     \errmessage{--> \noexpand\footnotetext command has no corresponding
          \noexpand\footnotemark}}

\def\err@labelredef#1{\immediate\write\sixt@@n
     {--> Label "#1" redefined}}

\def\err@badlabelmatch#1{\immediate\write\sixt@@n
     {--> Definition of label "#1" doesn't match value in \jobname.lab}}

\def\err@needlabel#1{\immediate\write\sixt@@n
     {--> Label "#1" cited before its definition}}

\def\err@undefinedlabel#1{\immediate\write\sixt@@n
     {--> Label "#1" cited but never defined}}

\def\err@undefinedeqn#1{\immediate\write\sixt@@n
     {--> Equation "#1" not defined}}

\def\err@undefinedref#1{\immediate\write\sixt@@n
     {--> Reference "#1" not defined}}

\def\err@nostream#1{%
     \errhelp={You have tried to input a stream file that doesn't exist.^^J}%
     \errmessage{--> Stream file #1 not found}}

\message{jyTeX initialization}

\everyjob{\immediate\write16{--> jyTeX version \fmtversion}%
     \edef\@@jobname{\jobname}%
     \edef\jobname{\@@jobname}%
     \settime
     \openin0=\jobname.lab
     \ifeof0
     \else\closein0
          \immediate\write16{--> Getting labels from file \jobname.lab}%
          \input\jobname.lab
     \fi}


\def\fixedskipslist{%
     \^^\{\topskip}%
     \^^\{\splittopskip}%
     \^^\{\maxdepth}%
     \^^\{\skip\topins}%
     \^^\{\skip\footins}%
     \^^\{\headskip}%
     \^^\{\footskip}}

\def\scalingskipslist{%
     \^^\{\p@renwd}%
     \^^\{\delimitershortfall}%
     \^^\{\nulldelimiterspace}%
     \^^\{\scriptspace}%
     \^^\{\jot}%
     \^^\{\normalbaselineskip}%
     \^^\{\normallineskip}%
     \^^\{\normallineskiplimit}%
     \^^\{\baselineskip}%
     \^^\{\lineskip}%
     \^^\{\lineskiplimit}%
     \^^\{\bigskipamount}%
     \^^\{\medskipamount}%
     \^^\{\smallskipamount}%
     \^^\{\parskip}%
     \^^\{\parindent}%
     \^^\{\abovedisplayskip}%
     \^^\{\belowdisplayskip}%
     \^^\{\abovedisplayshortskip}%
     \^^\{\belowdisplayshortskip}%
     \^^\{\abovechapterskip}%
     \^^\{\belowchapterskip}%
     \^^\{\abovesectionskip}%
     \^^\{\belowsectionskip}%
     \^^\{\abovesubsectionskip}%
     \^^\{\belowsubsectionskip}}


\def\twoupsetup{
     \topmargin=.75in
     \leftmargin=.5in
     \vsize=6.9in
     \hsize=4.75in
     \fullhsize=10in
     \let\draft=\relax}

\outputstyle{normal}                             

\def\marginnoteformat{\subscriptsize             
     \hsize=1in \baselinestretch=1000 \everypar={}%
     \tolerance=5000 \hbadness=5000 \parskip=0pt \parindent=0pt
     \leftskip=0pt \rightskip=0pt \raggedright}

\head={\ifdraft\normalfonts\it\hfil DRAFT\hfil   
     \llap{\number\day\ \monthword\month\ \militarytime}\else\hfil\fi}
\foot={\hfil\normalfonts\numstyle\pagenum\hfil}  

\normalbaselineskip=12pt                         
\normallineskip=0pt                              
\normallineskiplimit=0pt                         
\normalbaselines                                 

\topskip=.85\baselineskip \splittopskip=\topskip \headskip=2\baselineskip
\footskip=\headskip

\pagenumstyle{arabic}                            

\parskip=0pt                                     
\parindent=20pt                                  

\baselinestretch=1000                            


\chapterstyle{left}                              
\chapternumstyle{blank}                          
\def\chapterbreak{\newpage}                      
\abovechapterskip=0pt                            
\belowchapterskip=1.5\baselineskip               
     plus.38\baselineskip minus.38\baselineskip
\def\chapternumformat{\numstyle\chapternum.}     

\sectionstyle{left}                              
\sectionnumstyle{blank}                          
\def\sectionbreak{\vskip0pt plus4\baselineskip\penalty-100
     \vskip0pt plus-4\baselineskip}              
\abovesectionskip=1.5\baselineskip               
     plus.38\baselineskip minus.38\baselineskip
\belowsectionskip=\the\baselineskip              
     plus.25\baselineskip minus.25\baselineskip
\def\sectionnumformat{
     \ifblank\chapternumstyle\then\else\numstyle\chapternum.\fi
     \numstyle\sectionnum.}

\subsectionstyle{left}                           
\subsectionnumstyle{blank}                       
\def\subsectionbreak{\vskip0pt plus4\baselineskip\penalty-100
     \vskip0pt plus-4\baselineskip}              
\abovesubsectionskip=\the\baselineskip           
     plus.25\baselineskip minus.25\baselineskip
\belowsubsectionskip=.75\baselineskip            
     plus.19\baselineskip minus.19\baselineskip
\def\subsectionnumformat{
     \ifblank\chapternumstyle\then\else\numstyle\chapternum.\fi
     \ifblank\sectionnumstyle\then\else\numstyle\sectionnum.\fi
     \numstyle\subsectionnum.}


\footnotenumstyle{symbols}                       
\footnoteskip=0pt                                
\def\footnotenumformat{\numstyle\footnotenum}    
\def\footnoteformat{\footnotesize                
     \everypar={}\parskip=0pt \parfillskip=0pt plus1fil
     \leftskip=1em \rightskip=0pt
     \spaceskip=0pt \xspaceskip=0pt
     \def\\{\ifhmode\ifnum\lastpenalty=-10000
          \else\hfil\penalty-10000 \fi\fi\ignorespaces}}


\def\undefinedlabelformat{$\bullet$}             


\equationnumstyle{arabic}                        
\subequationnumstyle{blank}                      
\figurenumstyle{arabic}                          
\subfigurenumstyle{blank}                        
\tablenumstyle{arabic}                           
\subtablenumstyle{blank}                         

\eqnseriesstyle{alphabetic}                      
\figseriesstyle{alphabetic}                      
\tblseriesstyle{alphabetic}                      

\def\puteqnformat{\hbox{
     \ifblank\chapternumstyle\then\else\numstyle\chapternum.\fi
     \ifblank\sectionnumstyle\then\else\numstyle\sectionnum.\fi
     \ifblank\subsectionnumstyle\then\else\numstyle\subsectionnum.\fi
     \numstyle\equationnum
     \numstyle\subequationnum}}
\def\putfigformat{\hbox{
     \ifblank\chapternumstyle\then\else\numstyle\chapternum.\fi
     \ifblank\sectionnumstyle\then\else\numstyle\sectionnum.\fi
     \ifblank\subsectionnumstyle\then\else\numstyle\subsectionnum.\fi
     \numstyle\figurenum
     \numstyle\subfigurenum}}
\def\puttblformat{\hbox{
     \ifblank\chapternumstyle\then\else\numstyle\chapternum.\fi
     \ifblank\sectionnumstyle\then\else\numstyle\sectionnum.\fi
     \ifblank\subsectionnumstyle\then\else\numstyle\subsectionnum.\fi
     \numstyle\tablenum
     \numstyle\subtablenum}}


\referencestyle{sequential}                      
\referencenumstyle{arabic}                       
\def\putrefformat{\numstyle\referencenum}        
\def\referencenumformat{\numstyle\referencenum.} 
\def\putreferenceformat{
     \everypar={\hangindent=1em \hangafter=1 }%
     \def\\{\hfil\break\null\hskip-1em \ignorespaces}%
     \leftskip=\refnumindent\parindent=0pt \interlinepenalty=1000 }


\normalsize


\def\fmtversion{2.6M (June 1992)}

\catcode`\@=12

\typesize=10pt \magnification=1200 \baselineskip17truept
\footnotenumstyle{arabic} \hsize=6truein\vsize=8.5truein
\input epsf
\sectionnumstyle{blank}
\chapternumstyle{blank}
\chapternum=1
\sectionnum=1
\pagenum=0

\def\begintitle{\pagenumstyle{blank}\parindent=0pt
\begin{narrow}[0.4in]}
\def\endtitle{\end{narrow}\newpage\pagenumstyle{arabic}}


\def\beginexercise{\vskip 20truept\parindent=0pt\begin{narrow}[10
truept]}
\def\endexercise{\vskip 10truept\end{narrow}}


\def\eql#1{\eqno\eqnlabel{#1}}
\def\ref{\reference}
\def\peq{\puteqn}
\def\pref{\putref}

\def\mgn{\marginnote}
\def\bex{\begin{exercise}}
\def\eex{\end{exercise}}


\font\open=msbm10 

 
\def\StretchRtArr#1{{\count255=0\loop\relbar\joinrel\advance\count255 by1
\ifnum\count255<#1\repeat\rightarrow}}
\def\StretchLtArr#1{\,{\leftarrow\!\!\count255=0\loop\relbar
\joinrel\advance\count255 by1\ifnum\count255<#1\repeat}}

\def\StretchLRtArr#1{\,{\leftarrow\!\!\count255=0\loop\relbar\joinrel\advance
\count255 by1\ifnum\count255<#1\repeat\rightarrow\,\,}}

\def\mbox#1{{\leavevmode\hbox{#1}}}

\def\hspace#1{{\phantom{\mbox#1}}}
\def\oZ{\mbox{\open\char90}}


\def\Ga{\Gamma}

\def\th{\theta}

\def\ze{\zeta}

\def\De{\Delta}

\def\sc{{\rm sc }}

\def\zf{$\zeta$--function}


\def\frac#1/#2{\leavevmode\kern.1em
\raise.5ex\hbox{\the\scriptfont0 #1}\kern-.1em/\kern-.15em
\lower.25ex\hbox{\the\scriptfont0 #2}}
\def\sfrac#1/#2{\leavevmode\kern.1em
\raise.5ex\hbox{\the\scriptscriptfont0 #1}\kern-.1em/\kern-.15em
\lower.25ex\hbox{\the\scriptscriptfont0 #2}}

\def\gtorder{\mathrel{\raise.3ex\hbox{$>$}\mkern-14mu
             \lower0.6ex\hbox{$\sim$}}}
\def\ltorder{\mathrel{\raise.3ex\hbox{$<$}\mkern-14mu
             \lower0.6ex\hbox{$\sim$}}}

\def\semidirprod{\rlap{\ss C}\raise1pt\hbox{$\mkern.75mu\times$}}
\def\for{\lower6pt\hbox{$\Big|$}}
\def\fish{\kern-.25em{\phantom{abcde}\over \phantom{abcde}}\kern-.25em}


\def\boxit#1{\vbox{\hrule\hbox{\vrule\kern3pt
        \vbox{\kern3pt#1\kern3pt}\kern3pt\vrule}\hrule}}
\def\dalemb#1#2{{\vbox{\hrule height .#2pt
        \hbox{\vrule width.#2pt height#1pt \kern#1pt \vrule
                width.#2pt} \hrule height.#2pt}}}

\def\frac#1#2{{{#1}\over{#2}}}

\def\noin{\noindent}


\def\cosec{{\rm cosec\,}}

\def\eg{{\it e.g.}}

\def\pa{\partial}



\def\3j#1#2#3#4#5#6{\left\lgroup\matrix{#1&#2&#3\cr#4&#5&#6\cr}
\right\rgroup}

\def\m?{\mgn{?}}

\def\pa{\partial}

\def\beq{\begin{eqnarray}}
\def\eeq{\end{eqnarray}}


\def\aop#1#2#3{{\it Ann. Phys.} {\bf {#1}} ({#2}) #3}
\def\cjp#1#2#3{{\it Can. J. Phys.} {\bf {#1}} ({#2}) #3}
\def\cmp#1#2#3{{\it Comm. Math. Phys.} {\bf {#1}} ({#2}) #3}
\def\cqg#1#2#3{{\it Class. Quant. Grav.} {\bf {#1}} ({#2}) #3}

\def\jmp#1#2#3{{\it J. Math. Phys.} {\bf {#1}} ({#2}) #3}
\def\jpa#1#2#3{{\it J. Phys.} {\bf A{#1}} ({#2}) #3}

\def\np#1#2#3{{\it Nucl. Phys.} {\bf B{#1}} ({#2}) #3}

\def\pl#1#2#3{{\it Phys. Lett.} {\bf {#1}} ({#2}) #3}

\def\prp#1#2#3{{\it Phys. Rep.} {\bf {#1}} ({#2}) #3}

\def\prB#1#2#3{{\it Phys. Rev.} {\bf B{#1}} ({#2}) #3}
\def\prD#1#2#3{{\it Phys. Rev.} {\bf D{#1}} ({#2}) #3}
\def\prl#1#2#3{{\it Phys. Rev. Lett.} {\bf #1} ({#2}) #3}

\def\am#1#2#3{{\it Acta Mathematica} {\bf {#1}} ({#2}) #3}
\def\aim#1#2#3{{\it Adv. in Math.} {\bf {#1}} ({#2}) #3}
\def\ajm#1#2#3{{\it Am. J. Math.} {\bf {#1}} ({#2}) #3}

\def\jpamt#1#2#3{{\it J. Phys.A:Math.Theor.} {\bf{#1}} ({#2}) #3}
\def\jram#1#2#3{{\it J. f. reine u. Angew. Math.} {\bf {#1}} ({#2}) #3}

\def\ma#1#2#3{{\it Math. Ann.} {\bf {#1}} ({#2}) #3}

\def\mz#1#2#3{{\it Math. Zeit.} {\bf {#1}} ({#2}) #3}

\def\plb#1#2#3{{\it Phys. Letts.} {\bf {B#1}} ({#2}) #3}

\def\qjm#1#2#3{{\it Quart. J. Math.} {\bf {#1}} ({#2}) #3}

\def\rmjm#1#2#3{{\it Rocky Mountain J. Math.} {\bf {#1}} ({#2}) #3}

\def\tams#1#2#3{{\it Trans.Am.Math.Soc.} {\bf {#1}} ({#2}) #3}

\begin{title}
\vglue 0.5truein
\vskip15truept
\centertext {\Bigfonts \bf On sums of powers of cosecs} \vskip7truept
\centertext{\Bigfonts \bf } \vskip10truept\centertext{\Bigfonts \bf }
\centertext{J.S.Dowker\footnote{dowker@man.ac.uk;  dowkeruk@yahoo.co.uk}} \vskip
7truept \centertext{\it Theory Group,} \centertext{\it School of Physics and Astronomy,}
\centertext{\it The University of Manchester,} \centertext{\it Manchester, England} \vskip
7truept \centertext{}

 \begin{narrow} The finite sums of powers of cosecs occur in numerous
situations, both physical and mathematical, examples being the Casimir effect, R\'enyi
entropy, Verlinde's formula and Dedekind sums. I here present some further discussion
which consists mainly of a reprise of early work by H.M.Jeffery in 1862-64 which has fallen
by the wayside and whose results are being reproduced up to the present day. The
motivation is partly historical justice and partly that, because of the continuing
appearance of the sums, his particular methods deserve re--exposure. For example, simple
trigonometric generating functions are found and these have a field theoretic, Green
function significance and I make a few comments in the topic of R\'enyi entropies.

\end{narrow}
\end{title}

\section{\bf1. Introduction}
The finite summations of powers of cosecs occur in a number of different areas which,
mathematically, are associated with image constructions, in one way or another, and
involve the regular subdivisions or discretisations of the circle, a very old topic.

Relatively recent papers that detail some of the history behind these, and like
trigonometric summations, are Berndt and Yeap, [\pref{BandY}],  and Cvijovi\'c and
Srivastava, [\pref{CandS2,CandS}] and I refer to them for motivation, both physical and
mathematical. I will not repeat the references in these useful works, unless they are
directly relevant, but I might add a few more. These summations continue to appear in the
physics literature.

My intention here  to give some historical and calculational details which I hope will be
interesting and/or useful. I have elaborated the algebra because the main reference I
employ (Jeffery, [\pref{Jeffery}]) is, perhaps, a little obscure and certainly unknown. Also
the often quite simple methods have application today. I have also included an example
that is not in [\pref{Jeffery}] as a slight novelty as it combines bosonic and fermionic
elements.

The next section introduces the sums in their general form and a (known) computable
answer. Section 3 starts with Jeffery's generating function for the simplest, untwisted sum
for which an explicit expression is thence found.
\section{\bf2. The summations }
The summations in question are the modulated cosec sums,
  $$
  C_\nu(n,w)\equiv\sum_{l=1}^{n-1}\cos\bigg({2\pi w l\over n}\bigg)
  \,\cosec^{2\nu}\bigg({\pi l\over n}\bigg)\,,\quad \nu\in\oZ,
  \eql{snu}
  $$
with twisting $w\in\oZ$, In fact here I will deal mostly with the untwisted case $w=0$

The sums were evaluated in terms of generalised Bernoulli polynomials  by a contour
method, [\pref{Dowcascone}] and [\pref{Dow7}],
$$\eqalign{
  C_\nu(n,w)=
  {2^{2\nu}\over (2\nu)!}\,\,B^{(2\nu+1)}_{2\nu}(w+\nu\mid n,{\bf1})\,,\cr
  }
  \eql{res6}
  $$
which can be expanded as a polynomial in $n$.

This could be taken as the final answer, and it is,  but it is a little awkward to expand
hence I now pass to another expression, by a different approach, which yields a form
amenable to hand calculation.

\section{\bf3. Jeffery's generating function}

In fairly recent times generating functions for the (untwisted) summations have been
given by Fisher, [\pref{Fisher}], and Zagier, [\pref{Zagier}],  (and, more generally, for the
twisted ones in [\pref{Dow7}]) but the main point I wish to make in the present paper is
that a much earlier version was derived by Jeffery, [\pref{Jeffery}], in his discussion of
certain classes of integrals associated with the derivatives of the Gamma function.

Amongst other things, Jeffery derives the following result in \S36, \footnote{ I enlarge on
his algebra. Be aware that there are a number of bad misprints in this tersely written,
disjointed, but very interesting paper.}
  $$\eqalign{
  {\pi\over2n}&\sum_{l=1}^{n-1}\bigg(\cot{\pi\over n}(l-x)-\cot{\pi\over n}(l+x)\bigg)=
  \int_0^1 dy\,{1+y+\ldots+y^{n-2}\over1-y^n}\big(y^{-x}-y^x\big)\cr
  =&\int_0^1dy\,\bigg({1\over1-y}-{y^{n-1}\over1-y^n}\bigg)\big(y^{-x}-y^x\big)=
  \int_0^1dy\,\bigg({y^{-x}-y^x\over1-y}-{y^{-x/n}-y^{x/n}\over n(1-y)}\bigg)\cr
  =&2\bigg(1-{1\over n^2}\bigg)\,\ze(2)\,x+2\bigg(1-{1\over n^4}\bigg)\,\ze(4)\,x^3+\ldots\cr
  =&{\pi\over n}\cot{\pi x\over n}-\pi\cot\pi x\,,
  }
  \eql{genf}
  $$
where $n$ can be even or odd.

The third line results from expanding $y^x$ and then using,
$$\eqalign{
\int_0^1dt\,{\log^q t\over 1-t}&=\sum_{p=0}^\infty \int_0^1dt\,t^q\log^p t=(-1)^q\,
q!\sum_{p=0}^\infty{1\over (p+1)^{q+1}}\cr
&=(-1)^qq!\,\ze(q+1)\,.
}
$$

 I amplify Jeffery's derivation of the first line of the identity, (\peq{genf}), which he
does not give in detail. I will assume that any formula in (\pref{EMOT}) is known,
although Jeffery derives many of them anew. Substitution of the integral,
  $$
 \psi(z)=\int_0^1dt\,{1-t^{z-1}\over1-t}+\Ga'(1)\,,
  $$
into the standard result,
  $$
-\pi\cot \pi z=\psi(z)-\psi(1-z)\,
  $$
gives
  $$
-\pi\cot\pi z=\int_0^1dt\,{-t^{z-1}+t^{-z}\over1-t}\,.
\eql{cotrep}
  $$
Therefore, $z=(l\pm x)/n$,

$$\eqalign{
  &\pi\cot{\pi\over n}(l-x)-\pi\cot{\pi\over n}(l+x)\cr
  &=\int_0^1dt\,{-t^{(l+x)/n-1}+t^{-(l+x)/n}+t^{(l-x)/n-1}-
  t^{-(l-x)/n}\over1-t}\cr
  &=n\int_0^1dy\,{(y^{n-l-1}+y^{l-1})(y^{-x}-y^x)\over1-y^n}\,,\cr
}
  $$
where $t=y^{-n}$,

Hence, writing out the sum at length,
$$
\pi\sum_{l=1}^{n-1}\bigg(\cot{\pi\over n}(l-x)-\cot{\pi\over n}(l+x)\bigg)=
 2n \int_0^1 dy\,{1+y+\ldots+y^{n-2}\over1-y^n}\big(y^{-x}-y^x\big)\
$$
which is the first line in (\peq{genf}).

Equation (\peq{genf}) constitutes a generating function and is used as such by Jeffery to
evaluate the sums $C_\nu(m,0)$ by finding the (odd) derivatives with respect to $x$ at 0.

One might as well take the first derivative at once to give,

  $$\eqalign{
  {1\over2}\bigg({\pi\over n}\bigg)^{\!2}\,\,&\sum_{l=1}^{n-1}\bigg(\cosec^2
  {\pi\over n}(l-x)+\cosec^2{\pi\over n}(l+x)\bigg)\cr
  &=2\bigg(1-{1\over n^2}\bigg)\,\ze(2)+2.3\bigg(1-{1\over n^4}\bigg)\,\ze(4)\,x^2
  +2.5\bigg(1-{1\over n^6}\bigg)\,\ze(6)\,x^4+\ldots\cr
  &=\bigg(\pi^2\,\cosec^2\pi x-{\pi^2\over n^2}\,\cosec^2{\pi x\over n}\bigg)\,.
  }
  \eql{genf2}
  $$

Setting $x$ to zero yields the easiest sum,
  $$
  C_1(n,0)={1\over2}\sum_{l=1}^{n-1}\cosec^2\bigg({l\pi\over n}\bigg)
  ={n^2-1\over\pi^2}\,\ze(2)
  ={n^2-1\over6}\,.
  $$
Jeffery's paper is the earliest to which I can trace this result.

Equation (\peq{genf2}) is used in the following way. The sums $C_2(n,0)$ and $C_4(n,0)$
are given explicitly in [\pref{Jeffery}] as further examples and the construction of the
numerical coefficients exhibited there indicates that [\pref{Jeffery}] has employed,
without comment, the recursion formula, \footnote{ It is interesting to note that this
recursion is an early manifestation of Hadamard's technique,  of increasing the dimension
of the manifold by two by applying an intertwining operator to a propagation quantity.
Hadamard worked in flat space. An application to spheres was made by ourselves,
[\pref{DandC2}], and in a relevant conical context in [\pref{Dow6}]. See also
[\pref{Camporesi}].}
  $$
  \cosec^\nu y={1\over(\nu-1)(\nu-2)}\big(D_z^2+(\nu-2)^2\big)\,\cosec^{\nu-2}(y+z)
  \bigg|_{z=0}\,,\quad \nu>2\,,
  \eql{Ely1}
  $$
for even $\nu$, and $D_z=\pa/\pa z$, which must have been common knowledge, probably
going back to Euler. Much later references are Ely, [\pref{Ely}], and Saalsch\"utz,
[\pref{Saalschutz}].

The complete iteration of this  yields,
  $$\eqalign{
  \cosec^{2p} y&={1\over(2p-1)!}\big(D_z^2+(2p-2)^2\big)\ldots\big(D^2_z
  +2^2\big)\,\cosec^2(y+z)\bigg|_{z=0}\cr
  &={1\over(2p-1)!}\sum_{i=0}^{p-1}U^p_i\,D_z^{2i}\,\cosec^{2}(y+z)\bigg|_{z=0}\,,
  }
  \eql{iter}
  $$
so that equation (\peq{genf2}) can now be employed to give the required even derivatives
at zero,
  $$
  D^{2i}_z\sum_{l=1}^{n-1}\cosec^{\!2}\bigg({\pi l\over n}\pm z\bigg)\bigg|_{z=0}
  =2\,(2i+1)!\,\bigg({n^{2i+2}-1}\bigg)\,{\ze(2i+2)\over\pi^{2i+2}}\,.
  $$

Therefore, finally, setting $y=\pi l/n$ in (\peq{iter}) and summing,
  $$
  {1\over2}\sum_{l=1}^{n-1}\cosec^{\!2p}\bigg({\pi l\over n}\bigg)={2^{2p-2}\over(2p-1)!}
  \sum_{i=0}^{p-1}{(2i+1)!\over 2^{2i}}\,W_i^p\bigg({n^{2i+2}-1}\bigg)\,{\ze(2i+2)\over\pi^{2i+2}}\,,
  \eql{ans}
  $$
which I will refer to as Jeffery's form. The constants $U_i^p$ equal the sum of the
products of the squared first $p-1$ integers (excluding zero) taken $p-1-i$ at a time, but,
for later numerical convenience, I have extracted a factor of a power of two to give the
more usual constants, $W_i^p$.

A quite recent paper, [\pref{GandB}], has again considered these sums and reaches the
same expression as (\peq{ans}) by a similar method. It appears also in a recent work,
[\pref{GandP}].

\section{\bf4. The calculation of the coefficients}

The combinatorial expression for the coefficients, $U_i^p$, directly applied, is satisfactory
for a numerical hand calculation of the lower summations (Jeffery stops at $p=4$) but is
arduous for high $p$. For this reason consider the product occurring in (\peq{iter}). It is
clear that the coefficient of $D^{2i}$ equals $2^{2p-2-2i}$ times the coefficient of
$x^{2i}$, up to a sign, in the product

  $$\eqalign{
  \big(x^2-(p-1)^2\big)\ldots\big(x^2-1\big)=\,x^{[2p]-2}
  }
  $$
where $x^{[2p]}$ is a even central factorial, see \eg\ Steffensen, [\pref{Steffensen}]. I
have called this coefficient, $W_i^p$.

The expansion of the central factorial in terms of the central factorial coefficients (essentially
just a Taylor expansion),
  $$
  x^{{[2p]}-2}=\sum_{\nu=1}^p{D^{2\nu}\,0^{[2p]}\over(2\nu)!}\,x^{2\nu-2}\,,
  $$
is here expressed as central derivatives of nothing. These satisfy a recursion which allows
machine calculation. This is, [\pref{Steffensen}],
  $$
  {D^{2\nu}\,0^{[2p+2]}\over(2\nu)!}={D^{2\nu-2}\,0^{[2p]}\over(2\nu-2)!}
  -p^2\,{D^{2\nu}\,0^{[2p]}\over(2\nu)!}
  $$
with the starting values
  $$
  {D^{2}\,0^{[2p]}\over(2)!}=(-1)^{p-1}\big((k-1)!\big)^2\,;\quad
  {D^{2p}\,0^{[2p]}\over(2p)!}=1\,.
  $$

Hence the coefficients in (\peq{iter}) are, \footnote{ Jeffery, [\pref{Jeffery3}], gives a
symbolic form for the combinatorial description but it is expressed in terms of forward
differences and is more complicated.}
  $$
  W_i^p=(-1)^{p-1-i}\,{D^{2\nu}\,0^{[2p]}\over(2\nu)!}\,.
  \eql{cfn}
  $$

Early tabulations can be found in Steffensen, [\pref{Steffensen}], and Thiele,
[\pref{Thiele}] p.35 sufficient to reach $p=8$. Later tabulations exist (\eg\ in
[\pref{BSSV}]). Table 1 in the more recent discussion of these cosec sums by Grabner and
Prodinger, [\pref{GandP}], is the same as the one in Thiele and equivalent to that in
[\pref{GandB}].

I table some values taken from Thiele,
  $$
\matrix{p&1&2&3&4&5&i\cr{}&1&1&4&36&576&0\cr {}&{}&{1}&5&49&820&1
\cr{}&{}&{}&1&14&273&2\cr{}&{}&{}&{}&1&30&3\cr{}&{}&{}&{}&{}&1&{\,\,4}\,.
}
$$

Substitution of these numbers into the answer, (\peq{ans}), produces agreement with
existing results and, of course, with (\peq{res6}). There is no need to write out any
specific cases.

\section{\bf 5. Jeffery's  twisted generating function}

Jeffery also gives expressions for the (fermionic) summations,\footnote{ I give with
elaborations the, sometimes elementary, algebra because the reference may not be
available. The journal can actually be found in the {\it G\"ottinger
Digitalisierungscentrum}.}
  $$\eqalign{
  &-C_{m+1}\big(n,{n-1\over2}\big)=
  \sum_{l=1}^{n-1}(-1)^l\cos \bigg({\pi l\over n}\bigg)\,
  \cosec^{2m+2}\bigg({\pi l\over n}\bigg)\cr
  &=
  \sum_{l=1}^{n-1} (-1)^l\cot \bigg({\pi l\over n}\bigg)\,
  \cosec^{2m+1}\bigg({\pi l\over n}\bigg)\,,
  }
  \eql{defs}
  $$
which, again, he derives from a generating function,
   $$
   {\pi\over2n}\sum_{l=1}^{n-1}(-1)^{l+1}\bigg(\cosec{\pi\over n}(l-x)-
   \cosec{\pi\over n}(l+x)\bigg)=\pi\,\,\cosec\pi x-{\pi\over n}\cosec{\pi x\over n}\,,
   \eql{genff}
   $$
if $n$ is odd.

The proof proceeds as for the untwisted case and begins with the representation,
    $$
  \pi\,\cosec \pi x=\int_0^1dt\,{t^{x-1}+t^{-x}\over t+1}={1\over x}+\int_0^1
  dt\,{t^{x}-t^{-x}\over t+1}\,,\quad |x|<1\,,
  \eql{repn}
  $$
which follows by differentiating Kummer's integral, (re--proved by Jeffery),
  $$
\log\tan{\pi x\over 2}=\int_0^1dy\,\,{y^{x-1}-y^{-x}\over(1+y)\log y}\,.
  $$

Hence,
  $$
  \pi\,\cosec {\pi\over n}(l\mp x)
  =\int_0^1 dt\,{t^{{(l\mp x)}/n-1}-t^{-(l\mp x)/n}\over t+1}\,,
  \quad |x|<1\,,
  $$
if $l$ lies between $0$ and $n$

Set $t=y^{-n}$ then the integral leads to,\mgn{check signs}
  $$\eqalign{
  &{\pi\over n}\bigg(\cosec{\pi\over n}(l-x)-
   \cosec{\pi\over n}(l+x)\bigg)=\cr
&- \int_0^1dy\,{y^{-l+ x-1+n}-y^{l- x-1}-
y^{-l- x-1+n}-y^{l+ x-1}\over1+y^n}\cr
&=\int_0^1dy\,{\big(y^{-l-1+n}-y^{l-1}\big)
\big(y^{-x}-y^x\big)\over1+y^n}\,.\cr
}
 \eql{int4}
  $$
This allows the summation to be done.

Now put in the values of $l=1,2,\ldots,n-1$ in (\peq{genff}) with the correct parity. The
numerator of (\peq{int4}) contains ($n$ is odd),
  $$
y^{n-2}-y^{n-3}+\ldots-y+1+1-y+y^2-\ldots+y^{n-2}\,,
  $$
so that (\peq{int4}) becomes,
  $$\eqalign{
&{\pi\over 2n}\sum_{l=1}^{n-1}(-1^{l+1}\bigg(\cosec{\pi\over n}(l-x)-
   \cosec{\pi\over n}(l+x)\bigg)=\cr
&=\int_0^1dy\,{(1-y+y^2-\ldots+y^{n-2})
\big(y^{-x}-y^x\big)\over1+y^n}\cr
&=\int_0^1dy\,\bigg({1\over1+y}-{y^{n-1}\over1+y^n}\bigg)
\big(y^{-x}-y^x\big)\cr
&=\int_0^1dy\,\bigg({y^{-x}-y^x\over1+y}-
{1\over n}{y^{-x/n}-y^{x/n}\over1+y}\bigg)\cr
&=\pi\,\cosec\pi x-{\pi\over n}\cosec{\pi x\over n}
}
 \eql{int5}
  $$
after using the second representation in (\peq{repn}) for $\cosec$. We have proved
(\peq{genff}).

The procedure is exactly as previously. The  power series of the right--hand side of
equation (\peq{genff}) is obtained by expanding $y^x$ and using,
 $$\eqalign{
\int_0^1dt\,{\log^q t\over 1+t}&=\sum_{p=0}^\infty \int_0^1dt\,(-1)^p\,
t^p\log^q t=(-1)^q\,
q!\sum_{p=0}^\infty(-1)^p{1\over (p+1)^{q+1}}\cr
&=(-1)^q\,q!\,\eta(q+1)\,,
}
$$
where $\eta$ is the Dirichlet $\eta$--function, related to the Riemann \zf\ by
$\eta(q+1)=\big(1-2^{-q}\big)\,\ze(q+1)$. (Note that the trigonometric closed form of
the generating function has not actually been used.)

This results in,
  $$
{\pi\over 2n}\sum_{l=1}^{n-1}(-1^{l+1}\bigg(\cosec{\pi\over n}(l-x)-
   \cosec{\pi\over n}(l+x)\bigg)=
2\,\sum_{j=1}^\infty\bigg(1-{1\over n^{2j}}\bigg)\,\eta(2j)\,x^{2j-1}\,,
  $$
which I differentiate with respect to $x$ to get,
  $$\eqalign{
&{1\over2}\bigg({\pi\over n}\bigg)^{\!\!2}\,
\sum_{l=1}^{n-1}(-1^{l+1}\bigg(\cot{\pi\over n}
(l-x)\,\cosec{\pi\over n}(l-x)+\cot{\pi\over n}(l+x)\,
   \cosec{\pi\over n}(l+x)\bigg)\cr
&=2\,\sum_{j=1}^\infty(2j-1)\bigg(1-{1\over n^{2j}}\bigg)\,\eta(2j)\,x^{2j-2}\,,
}
\eql{genfff}
  $$

For the left--hand side, note that
  $$
   D_z\cosec^{2m+1}(y+ z)=-(2m+1)\,\cosec^{2m+1}(y+x)\,\cot (y+z)
   \eql{drel}
  $$
and the iteration for $\cosec^{2m+1}\cot$ is  determined by that for $\cosec^{2m+1}$,
which is,
  $$
  \cosec^{2m+1}y=\pm{1\over(2m)!}\big(D_z^2+(2m-1)^2\big)
  \ldots \big(D_z^2+1\big)\,\cosec (y\pm z)\bigg|_{z=0}
  \eql{iter4}
  $$
and can again be expanded in terms of central factorial numbers. I write out the expansion
in the form,
  $$
 \cosec^{2m+1}y=\pm{1\over (2m)!}\sum_{i=0}^{m}V^m_i D_z^{2i}
 \cosec(y\pm z)\bigg|_{z=0}
 \eql{cosodd}
  $$

Differentiating (\peq{cosodd}) with respect to $y$, using (\peq{drel}), setting $y=\pi
l/n$, $z=\pm\pi x/n$ and taking the sum,
  $$\eqalign{
  &\sum_{l=1}^{n-1}(-1)^{l+1}\cot {\pi l\over n}
  \,\cosec^{2m+1}{\pi l\over n}\cr
  &={1\over (2m+1)!}
  \sum_{i=0}^{m}V_i^m\bigg({n\over\pi}\bigg)^{2i}
  D_x^{2i}\sum_{l=1}^{n-1}(-1)^{l+1}
  \bigg(\cot{\pi\over n}(l\pm x)\,\cosec{\pi\over n}(l\pm x)\bigg)\bigg|_{x=0}
  }
    $$

Substituting in the  derivatives at $x=0$, which  can be read off from (\peq{genff}),
delivers the final result, for $n$ odd,
  $$\eqalign{
  &\sum_{l=1}^{n-1}(-1)^{l+1}\cot {\pi l\over n}
  \,\cosec^{2m+1}{\pi l\over n}\cr
  &={2\over (2m+1)!}
  \sum_{i=0}^{m}(2i+1)!\,V_i^m\,\big(n^{2i+2}-1\big)\,
  {\eta(2i+2)\over\pi^{2i+2}}\,,
  }
  \eql{fr}
    $$
which should be compared with the bosonic sum, (\peq{ans}). One notices now the
appearance of the Dirichlet function, typical for fermionic quantities.

From the form of the full iteration, (\peq{iter4}), the coefficients, $V^m_i$, are given by
the sums of products of the squares of the {\it odd} natural numbers, and this is how
Jeffery, [\pref{Jeffery}] \S41, computes them. Another way is to employ a recursion
relation but the easiest option is to use Thiele, [\pref{Thiele}], p.36 who conveniently
tabulates them as the positive integers they are. These are sufficient to reach $m=8$.

For the convenience of the reader I lift a few lower values from [\pref{Thiele}],
$$
\matrix{m&0&1&2&3&4&i\cr{}&1&1&9&225&11025&0\cr {}&{}&1&10&259&12916&1
\cr{}&{}&{}&1&35&1974&2\cr{}&{}&{}&{}&1&84&3\cr{}&{}&{}&{}&{}&1&{\,\,4}\,.
}
$$

The tables in Steffensen, [\pref{Steffensen}], and in [\pref{BSSV}] can also be consulted.

As an example, Jeffery writes out the specific case $m=2$ and I here repeat it, putting in
the numerical values of the coefficients to give an explicit polynomial,
  $$
\sum_{l=1}^{n-1}(-1)^{l+1}\cot {\pi l\over n}
  \,\cosec^{\!\!5}{\pi l\over n}={1\over80}(n^2-1)+{7\over720}(n^4-1)
  +{31\over15120}(n^6-1)\,,
  $$
if $n$ is odd.

\section{\bf6. A mixed summation of alternating cosecs}
To ring the changes I compute, using Jeffery's method, the alternating sum
  $$
  -C_\nu(n,n/2)\equiv\sum_{l=1}^{n-1}(-1)^{l+1}
  \,\cosec^{2\nu}\bigg({\pi l\over n}\bigg)\,,\quad \nu\in\oZ,
  \eql{snu2}
  $$
for even $n$, which is neither bosonic nor fermionic. The algebra is a mixture of that in
sections 3 and 5 and starts with the representation (\peq{cotrep})  for the cotan, so that
$$\eqalign{
  {\pi\over2n}&\sum_{l=1}^{n-1}(-1)^{l+1}\bigg(\cot{\pi\over n}(l-x)-\cot{\pi\over n}(l+x)\bigg)\cr
  &=
  \int_0^1 dy\,{1-y+y^2-\ldots+y^{n-2}\over1-y^n}\big(y^{-x}-y^x\big)\cr
  =&\int_0^1dy\,\bigg({1\over1+y}+{y^{n-1}\over1-y^n}\bigg)\big(y^{-x}-y^x\big)=
  \int_0^1dy\,\bigg({y^{-x}-y^x\over1+y}+{y^{-x/n}-y^{x/n}\over n(1-y)}\bigg)\cr
  =\,&2\bigg(\eta(2)+{\ze(2)\over n^2}\bigg)\,x+2\bigg(\eta(4)+{\ze(4)
  \over n^4}\bigg)\,x^3
  +\ldots\cr
  =\,&\pi\,\cosec\pi x{\pi\over n}-\cot{\pi x\over n}\,.
  }
  \eql{fgen}
  $$

Differentiating with respect to $x$,
  $$\eqalign{
  {1\over2}&\bigg({\pi\over n}\bigg)^{\!2}\,\,\sum_{l=1}^{n-1}(-1)^{l+1}\bigg(\cosec^{\!2}
  {\pi\over n}(l-x)+\cosec^{\!2}{\pi\over n}(l+x)\bigg)\cr
  &=2\bigg(\eta(2)+{1\over n^2}\ze(2)\bigg)+2.3\bigg(\eta(4)+{1\over n^4}\ze(4)
  \bigg)\,x^2
  +2.5\bigg(\eta(6)+{1\over n^6}\ze(6)\bigg)\,x^4+\ldots
  }
  \eql{genf3}
  $$

Again, the recursion for $\cosec^{2p}x$ is brought in and gives, as before, for $n$ even,

$$\eqalign{
  &{1\over2}\sum_{l=1}^{n-1}(-1)^{l+1}
  \,\cosec^{2p}{\pi l\over n}\cr
  &={2^{2p-2}\over (2p+1)!}
  \sum_{i=0}^{p}{(2i+1)!\over2^{2i}}\,W_i^p\,\bigg(n^{2i+2}\eta(2i+2)+\ze(2i+2)\bigg)\,
  {1\over\pi^{2i+2}}\,,\cr
  }
  \eql{fb}
    $$
which has both bosonic and fermionic aspects. The purely fermionic summation,
(\peq{defs}), contains an extra `phase factor' which is just the trace of the spin-1/2
rotation matrix through angle $2\pi/n$ which is necessary because of the rotation of the
zwei-beine, [\pref{Dowcos}].

Formula (\peq{fb}) is easily programmed. Numerical examples agree with the general
expression, (\peq{res6}). Chu and Marini have computed some examples, [\pref{CandM}]
p.149, using a method somewhat less convenient than the one employed here, and I have
found agreement.

\begin{ignore}
\section{\bf 7. Application to field theory}

In this section a further use for the generating functions is presented. In fact they have
not really been used in the preceding, only the power series expansions are needed, and
these follow at a prior stage.

The situation occurs in the general area of entanglement entropy and conformal field
theory where free fields are often used as a more or less soluble case and I refer,
specifically, to the papers of Cardy, [\pref{Cardy}], and Herzog and Nian, [\pref{HandN}].
I do not give any background but, for convenience, start immediately with the free scalar
as expounded in [\pref{HandN}]. There the Green function in $d$ dimensions is given as
\footnote{ I will not give any earlier physical references where this standard quantity
occurs.}
  $$
  G^B_{1/n,d}(2\th)={1\over4}\sum_{l=0}^{n-1}\cosec^{d-2}\bigg(\th+{\pi l\over n}\bigg)\,.
  $$

To agree with the notation in the present paper, $n$ here is $m$ in
\end{ignore}

\section{\bf7. Comments, connections and conclusion}

As mentioned in the Introduction, the polynomials crop up in several seemingly disparate
topics, examples being statistical mechanics, vector bundle theory, [\pref{Dow7}],
Dedekind sums, interpolation, quantum field theory on multiply connected manifolds, such
as lens spaces, or on orbifolds with conical structures, real or artificial.

In the latter category is the construction of the entanglement and R\'enyi entropies using
the replica technique which introduces a conical singularity into a codimension 2
submanifold. This process is usually accomplished by using integer coverings but can
equally well be done with images as is well known. In this connection, Cardy,
[\pref{Cardy}], has also recently recalculated the classic summations, $C_1(n,0)$ and
$C_2(n,0)$. Furthermore in a related subject, Herzog and Nian, [\pref{HandN}], have
employed the recursion (\peq{Ely1}).

In this regard it is interesting to note that the simple rewriting of the generating function
(\peq{genf2}),\footnote{ This is a by now standard identity, [\pref{Bromwich}] p.211.}
$$\eqalign{
  \,&\sum_{l=0}^{n-1}
  \cosec^2{\pi\over n}(l\pm x)\cr
  &=n^2\,\cosec^2\pi x\,,
  }
  \eql{genf22}
  $$
gives the Green function in four dimensions as in [\pref{HandN}] equn.(33), to a constant
factor.

More general summations (in higher dimensions) of the same form appear in the appendix
to [\pref{Dow7}].

It should be remarked that, for the requirements of the present paper, the construction of
generating functions such as (\peq{genf}) is somewhat of an afterthought. Only the, prior
derived, power series are used.

The relation between coverings and images can be illustrated by Lubbock's summation
formulae, [\pref{Steffensen}], in interpolation theory as described in [\pref{dowlubb}]. All
I point out here is that the Lubbock approximation involves a series of polynomials which,
in one variant, are shown, [\pref{dowlubb}], by contours, to be given by,
  $$
  P_{2\nu}(n)={1\over(2\nu)!}{1\over n}\,B^{(2\nu+1)}_{2\nu}\big(\nu+(n-1)/2
  \mid n,{\bf 1}\big)\,,
  $$
in terms of generalised Bernoulli polynomials, or,
  $$
  P_{2\nu}(n)={1\over2^{2\nu}}{1\over n}\,C_\nu\big(n,(n-1)/2\big)\,,
  \eql{pee}
  $$
in terms of the fermionic summations.\footnote{ Another Lubbock variant gives
polynomials, $Q$, which are the untwisted sums.} Here, $n$ would have to be an integer
for the sum to make sense. After the summation has been effected, $n$ could be anything
and in the context of Lubbock's formula $n$ is set equal to $1/h$ with $h$ integral
providing the alternative form (see Steffensen, [\pref{Steffensen}]) in terms of central
factorials,
    $$
   P_{2\nu}(1/h)={1\over(2\nu)!}\,\sum_{-(h-1)/2}^{(h-1)/2}
   \bigg({j\over h}\bigg)^{[2\nu]}
    $$
where the sum is over $j$ in steps of one.

Steffensen lists a few examples of the $P$. These check with (\peq{pee}) which is no
surprise in view of the image relation,
$$
   \sum_{s=0}^{h-1} B^{(n+1)}_\nu\big(a+{s\over h}\mid {\bf 1}\big)=
   h\,B^{(n+1)}_\nu\big(a\mid {1\over h},{\bf 1}\big)\,.
   \eql{image2}
  $$
It might be helpful to spell things out. The left--hand side is the pre-image sum giving the
quantity (here $B$) on the (multiply connected) circle of circumference $1/h$ in terms of
that on a (bigger) covering \footnote{ The term `covering' is meant in the projection
sense.} circle of unit circumference consisting of $h$ copies of the smaller circle in the
form of circumference intervals of length $1/h$ with {\it interval} ends. This can be
referred to as a {\it subdivision} of the bigger circle.

By contrast, an integer $n$--cover of the unit circle means $n$ copies of the unrolled unit
circle joined end to end and the two boundary points identified giving a wrapped up circle
of circumference $h$. The left--hand side of the sum, (\peq{image2}), then does not make
sense but the right--hand side does.

When embedded in the plane, all circles have the same radius and give conical
singularities with positive and negative deficit angles for subdivisions and integer
coverings respectively.

The specific summations considered here all fall into the general twisted class,
(\peq{snu}), and if one is looking for just a numerical polynomial then the general
expression, (\peq{res6}), in terms of Bernoulli polynomials provides it. Jeffery's forms,
(\peq{ans}) and (\peq{fr}) (and (\peq{fb}) ) arrange these polynomials in a different,
more explicit manner. This arrangement has advantages when considering the heat--kernel
coefficients in the presence of conical singularities, [\pref{Dowker2}].

Everything that has been given can be transcribed by replacing $\cot$ by $\tan$ and
$\cosec$ by $\sec$ because $\sec$ and $\cosec$ obey the same recursion. (Some
adjustment of the summation is needed.)

\newpage
 \noin{\bf References.} \vskip5truept
\begin{putreferences}
  \ref{CandS}{}
  \ref{Bromwich}{Bromwich, T.J.I'A. {\it Infinite Series},
  (Macmillan, London, 1947).}
  \ref{Dowker2}{Dowker,J.S.\cqg{11}{1994}{L137}.}
  \ref{Camporesi}{Camporesi,R. \prp{196}{1990}{1}.}
  \ref{Dow6}{Dowker,J.S. \jmp{30}{1989}{770}.}
  \ref{DandC2}{Dowker,J.S. and Critchley,R. \prD{13}{1976}{224}.}
  \ref{CandM}{Chu,W. and Marini,A. {\it Advances in Appl.Math.} {\bf 23} (1999) 115.}
  \ref{Zagier}{Zagier,D., {\it Israel Math. Conf. Proc.} {\bf 9} (1996) 445.}
  \ref{Fisher}{Fisher,M. {SIAM Rev.} {\bf 13} (1969) 116 (solution of problem 69-14).}
  \ref{Cardy}{Cardy,J. \jpa{46}{2013} 285402.}
  \ref{Saalschutz}{Saalsch\"utz, L. \jram{126}{1903}{99}.}
  \ref{HandN}{Herzog,C.P. and Nian,J. {\it Thermal corrections to R\"enyi
  entropies for conformal field theories} ArXiv:1411.6505}
  \ref{dowlubb}{Dowker,J.S. {\it Remarks on Lubbock's summation formulae}
  ArXiv:1307.7067}
  \ref{EMOT}{Erdelyi, A., Magnus, W., Oberhettinger, F. and Tricomi, F.G. {
  \it Higher Transcendental Functions} Vol.1 (McGraw-Hill, N.Y. 1953).}
  \ref{Poinsot}{Poinsot,M.L. {\it Recherches sur l'Analyse des Sections Angulaires}
   (Bachelier, \break Paris, 1825).}
  \ref{Dowcen}{Dowker,J.S. {\it Central Differences, Euler numbers and
   symbolic methods} ArXiv: 1305.0500.}
   \ref{Dow7}{Dowker,J.S. \jpa{25}{1992}{2641}.}
  \ref{Dowcascone}{Dowker,J.S. \prD{36}{1987}{3095}.}
  \ref{Dowcos}{Dowker,J.S. \prD{36}{1987}{3742}.}
  \ref{Dowpower}{Dowker,J.S. {\it Poweroids revisited -- an old symbolic approach.}ArXiv:1307.3150.}
   \ref{Ely}{Ely,G.S. \ajm{5}{1882}{337}.}
  \ref{BSSV}{Butzer,P.L., Schmidt,M., Stark,E.L. and Vogt,I. {\it Numer.Funct.Anal.Optim.}
    {\bf 10} (1989) 419.}
  \ref{Steffensen}{Steffensen,J.F. {\it Interpolation}, (Williams and Wilkins,
    Baltimore, 1927).}
    \ref{BandY}{Berndt,B.C. and Yeap,B.P. {\it Adv. Appl. Math.}
  {\bf29} (2002) 358.}
    \ref{GandP}{Grabner,P.J. and Prodinger,H. {\it Quaestiones Math.} {\bf 30} (2007) 159 .}
    \ref{GandB}{Gauthier,N. and Bruckman,P.S. {\it Fibonacci Quart.} {\bf 44}  (2007) 264.}
    \ref{Thiele}{Thiele,T.N. {\it Interpolationsrechnung} (Teubner, Leipzig, 1909).}
\ref{CandS}{Cvijovi\'c,D. and Srivastava,H.M. \jpa{45}{2012}{374015}.}
   \ref{CandS2}{Cvijovi\'c,D. and Srivastava,H.M. \jmp{48}{2007}{043507}.}
    \ref{EandM}{Espinosa,O. and Moll,V.H. {\it The Ramanujan Journal} {\bf6}
(2002) 449.}
   \ref{Gosper}{Gosper,R,W.}
     \ref{Jeffery3}{Jeffery, H.M. \qjm{4}{1861}{364}.}
     \ref{BarnesGf}{Barnes,E.W.{\it Quart.J.Pure and Applied Maths.}
     {\it 31} (1899) 264.}
   \ref{Jeffery2}{Jeffery, H.M. \qjm{5}{1862}{91}.}
   \ref{Jeffery}{Jeffery, H.M. \qjm{6}{1864}{82}.}
 \ref{Norlund}{N\"orlund,N.E. \am{43}{1922}{121}.}
 \ref{Glaisher}{Glaisher,J.W.L. {\it Messenger of Math.} {\bf 7} (1878) 43.}
 \ref{Dow}{Dowker,J.S. {\it Poweroids revisited -- an old symbolic approach.}ArXiv:1307.3150.}
 \ref{KOW}{Kurokawa,N,, Ochiai,H. and Wakayama, M. {\it J.Ramanujan Math.Soc.} {\bf21}
 (2006) 153.}
 \ref{KandO}{Kurokawa,N. and Ochiai,H. {\it Kodaira Math.J.} {\bf30} (2007) 195.}
    \ref{Elizalde}{Elizalde,E. {\it Math. of Comp.} {\bf 47} (1986) 347.}
    \ref{kink}{Kinkelin,H. {\it J.f.reine u. angew. Math. (Crelle)} {\bf 57} (1860)
   122.}
  \ref{holder}{O.H\"older {\it G\"ott. Nachrichten} (1886) 514-522.}
  \ref{alex}{W.P.Alexeiewsky {\it Leipzig Berichte} {\bf 46} (1894) 268-275.}
     \ref{bend}{Bendersky,L. \am{61}{1933}{263}.}
     \ref{Adamchik1}{Adamchik, V.S.
   {\it J. Comp. Appl. Math.} {\bf 100} (1998) 191.}
   \ref{Adamchik}{V.S.Adamchik, {\it Contributions to the theory of the
   Barnes function}, ArXiv: math. CA/0308086.}
    \ref{FandT}{Fradkin, E.S and Tseytin,A.A. \pl{B134}{1984}{301}.}
     \ref{CEZ2}{Cognola,G.,Elizalde,E. and Zerbini,S.  \cmp{237}{2003}{507}.}
   \ref{Tseytlin2}{Tseytlin, A. \np{877}{2013}{632}.}
   \ref{Tseytlin}{Tseytlin,A.A. \np{877}{2013}{598}.}
  \ref{Dowma}{Dowker,J.S. {\it Calculation of the multiplicative anomaly} ArXiv: 1412.0549.}
  \ref{CandH}{Camporesi,R. and Higuchi,A. {\it J.Geom. and Physics}
  {\bf 15} (1994) 57.}
  \ref{Allen}{Allen,B. \np{226}{1983}{228}.}
  \ref{Dowdgjms}{Dowker,J.S. \jpamt{48}{2015}{125401}.}
  \ref{Dowsphgjms}{Dowker,J.S. {\it Numerical evaluation of spherical GJMS determinants
  for even dimensions}, ArXiv:1310.0759.}
  \ref{CEZ}{Cognola,G.,Elizalde,E. and Zerbini,S.  \jpamt{48}{2015}{045203}.}
 \ref{CFM}{Castillo-Garate,V.,Friedman,E. and M\v{a}ntoiu,M. {\it The multiplicative
 anomaly of three or more commuting elliptic operstors}, ArXiv:1211.4117.}
  \ref{DowGJMS}{Dowker,J.S.  \jpa{44}{2011}{115402}.}
  \ref{Dowcmp}{Dowker,J.S. \cmp{162}{1994}{633}.}
\ref{Dowbfe}{Dowker,J.S. {\it The boundary F-theorem for free fields}, ArXiv:1407.5909.}
    \ref{DandKi}{Dowker,J.S. and Kirsten, K. {\it Comm. in Anal. and Geom.}
    {\bf7} (1999) 641.}
         \ref{Kassel}{Kassel,C. Seminaire Bourbaki  n. 708 (1988-89) 199.}
              \ref{Vardi}{Vardi,I. {\it SIAM J.Math.Anal.} {\bf 19} (1988) 493.}

  \ref{CandT2}{Copeland,E. and Toms,D.J. \cqg {3}{1986}{431}.}
   \ref{DoandKi} {Dowker.J.S. and Kirsten, K. {\it Analysis and Appl.}
       {\bf 3} (2005) 45.}
\begin{ignore}

  \ref{CandT}{Copeland,E. and Toms,D.J. \np {255}{1985}{201}.}
  \ref{Apps}{Apps,J.S. Thesis (University of Manchester, 1995).}
 \ref{Allen2}{Allen,B. PhD Thesis, University of Cambridge, 1984.}
 \ref{Chodos1}{Chodos,A. and Myers,E. \aop{156}{1984}{412}.}
     \ref{Guillarmou}{Guillarmou,C. \ajm{131}{2009}{1359}.}

     \ref{BaandDu}{Basar,G and Dunne,G.V. \jpa {43}{2010}{072002}.}
     \ref{AaandD}{Aros,R. and Diaz,D.E. {\it Determinant and Weyl anomaly of
     Dirac operator: a holographic derivation}, ArXiv:1111.1463.}
     \ref{BaandS}{B\"ar,C. and Schopka,S. The Dirac determinant of spherical
     space forms,\break {\it Geom.Anal. and Nonlinear PDEs} (Springer, Berlin, 2003).}
     \ref{QandC}{J.R.Quine and J.Choi, \rmjm {26}{1996}{719-729}.}
  \ref{QandC2}{J.R.Quine and J.Choi, Zeta regularized products and functional
  determinants on spheres \rmjm {26}{1996}{719-729}.}
  \ref{moller}{M{\o}ller,N.M. \ma {343}{2009}{35}.}
     \ref{KKY}{Kanemitsu,S., Kumagai,H. and Yoshimoto,M. {\it The Ramanujan
    J.} {\bf 5}(2001)5.}
     \ref{KandB}{Kamela,M. and Burgess,C.P. \cjp{77}{1999}{85}.}
    \ref{Dowcos}{Dowker,J.S. \prD{36}{1987}{3742}.}
     \ref{Dow20}{Dowker,J.S. \jmp{35}{1994}{6076}.}
     \ref{DowGJMSO}{Dowker,J.S. {\it Numerical evaluation of spherical GJMS operators}
     ArXiv: \break 1309.2873.}
     \ref{Adamchik1}{V.S.Adamchik, Polygamma functions of negative order
   {\it J. Comp. Appl. Math.} {\bf 100} (1998) 191-199.}
   \ref{Adamchik}{V.S.Adamchik, {\it Contributions to the theory of the
   Barnes function}, ArXiv: math. CA/0308086.}

    \ref{Onodera}{Onodera,K. \aim{224}{2010}{895}.}
    \ref{KKY}{Kanemitsu,S., Kumagai,H. and Yoshimoto,M. {\it The Ramanujan
    J.} {\bf 5}(2001)5.}
     \ref{Juhl}{Juhl,A. {\it On conformally covariant powers of the Laplacian}
     ArXiv: 0905.3992}
     \ref{Elphinstone}{Elphinstone,H.W. \qjm {2}{1858}{252}.}
       \ref{Branson}{Branson,T.P. \tams{347} {1995}{3671}.}
      \ref{Bromwich}{Bromwich, T.J.I'A. {\it Infinite Series},
  (Macmillan, London, 1926).}
   \ref{Loney}{Loney, S.L. {\it Plane Trigonometry} (CUP, Cambridge, 1893).}
    \ref{Hertzberg}{Hertzberg,M.P. \jpa{46}{2013}{015402}.}
     \ref{CaandW}{Callan,C.G. and Wilczek,F. \plb{333}{1994}{55}.}
    \ref{CaandH}{Casini,H. and Huerta,M. \plb{694}{2010}{167}.}
    \ref{Lindelof}{Lindel\"of,E. {\it Le Calcul des Residues} (Gauthier--Villars, Paris,1904).}
    \ref{CaandC}{Calabrese,P. and Cardy,J. {\it J.Stat.Phys.} {\bf 0406} (2004) 002.}
    \ref{MFS}{Metlitski,M.A., Fuertes,C.A. and Sachdev,S. \prB{80}{2009}{115122}.}
    \ref{Gromes}{Gromes, D. \mz{94}{1966}{110}.}
    \ref{Pockels}{Pockels, F. {\it \"Uber die Differentialgleichung $\De
  u+k^2u=0$} (Teubner, Leipzig. 1891).}
   \ref{Diaz}{Diaz,D.E. JHEP {\bf 0807} (2008) 103.}
    \ref{DandD}{Diaz,D.E. and Dorn,H. JHEP {\bf 0705} (2007) 46.}
  \ref{Minak}{Minakshisundaram,S. {\it J. Ind. Math. Soc.} {\bf 13} (1949) 41.}
    \ref{CaandWe}{Candelas,P. and Weinberg,S. \np{237}{1984}{397}.}
     \ref{Chodos1}{Chodos,A. and Myers,E. \aop{156}{1984}{412}.}
     \ref{ChandD}{Chang,P. and Dowker,J.S. \np{395}{1993}{407}.}
    \ref{LMS}{Lewkowycz,A., Myers,R.C. and Smolkin,M. {\it Observations on
    entanglement entropy in massive QFTs.} ArXiv:1210.6858.}
    \ref{Bierens}{Bierens de Haan,D. {\it Nouvelles tables d'int\'egrales
  d\'efinies}, (P.Engels, Leiden, 1867).}

    \ref{Dowren}{Dowker,J.S. \jpamt {46}{2013}{2254}.}
    \ref{Doweven}{Dowker,J.S. {\it Entanglement entropy on even spheres.}
    ArXiv:1009.3854.}
     \ref{Dowodd}{Dowker,J.S. {\it Entanglement entropy on odd spheres.}
     ArXiv:1012.1548.}
    \ref{DeWitt}{DeWitt,B.S. {\it Quantum gravity: the new synthesis} in
    {\it General Relativity} edited by S.W.Hawking and W.Israel (CUP,Cambridge,1979).}
    \ref{Nielsen}{Nielsen,N. {\it Handbuch der Theorie von Gammafunktion}
    (Teubner,Leipzig,1906).}
    \ref{KPSS}{Klebanov,I.R., Pufu,S.S., Sachdev,S. and Saddi,B.R.
    {\it JHEP} 1204 (2012) 074.}
    \ref{KPS2}{Klebanov,I.R., Pufu,S.S. and Safdi,B.R. {\it F-Theorem without
    Supersymmetry} 1105.4598.}
    \ref{KNPS}{Klebanov,I.R., Nishioka,T, Pufu,S.S. and Safdi,B.R. {\it Is Renormalized
     Entanglement Entropy Stationary at RG Fixed Points?} 1207.3360.}
    \ref{Stern}{Stern,W. \jram {79}{1875}{67}.}
    \ref{Gregory}{Gregory, D.F. {\it Examples of the processes of the Differential
    and Integral Calculus} 2nd. Edn (Deighton,Cambridge,1847).}
    \ref{MyandS}{Myers,R.C. and Sinha, A. \prD{82}{2010}{046006}.}
   \ref{RyandT}{Ryu,S. and Takayanagi,T. JHEP {\bf 0608}(2006)045.}

     \ref{Dowjmp}{Dowker,J.S. \jmp{35}{1994}{4989}.}
      \ref{Dowhyp}{Dowker,J.S. \jpa{43}{2010}{445402}.}
       \ref{HandW}{Hertzberg,M.P. and Wilczek,F. \prl{106}{2011}{050404}.}
      \ref{dowkerfp}{Dowker,J.S.\prD{50}{1994}{6369}.}
       \ref{Fursaev}{Fursaev,D.V. \plb{334}{1994}{53}.}
       \ref{Barnesa}{Barnes,E.W. {\it Trans. Camb. Phil. Soc.} {\bf 19} (1903) 374.}
  \ref{Barnesb}{Barnes,E.W. {\it Trans. Camb. Phil. Soc.}{\bf 19} (1903) 426.}
  \end{ignore}
\end{putreferences}

 \bye